\documentclass[11pt,pre,superscriptaddress,bibnotes,showkeys]{revtex4}
\usepackage{epsfig}
\usepackage{times}
\newtheorem{defn}{Definition}[section]

\newtheorem{conj}[defn]{Conjecture}

\newtheorem{cor}[defn]{Corollary}

\newtheorem{lemma}[defn]{Lemma}

\newtheorem{thm}[defn]{Theorem}
\newtheorem{theorem}[defn]{Theorem}

\newcommand{\be}{\begin{equation}}
\newcommand{\ee}{\end{equation}}
\newcommand{\bea}{\begin{eqnarray}}
\newcommand{\eea}{\end{eqnarray}}
\newcommand{\beas}{\begin{eqnarray*}}
\newcommand{\eeas}{\end{eqnarray*}}
\newcommand{\goto}{\rightarrow}
\newcommand{\ink}{\rule{.5\baselineskip}{.55\baselineskip}}
\newcommand{\ds}{\displaystyle}
\newcommand{\ts}{\textstyle}
\newcommand{\noi}{\noindent}
\newcommand{\lan}{\langle}
\newcommand{\ran}{\rangle}

\newcommand{\skp}{\vspace{\baselineskip}}

\newcommand{\per}{\hspace{-.072in}{\bf .  }}

\newcommand{\R}{I\!\!R}
\renewcommand{\r}{I\!\!R}

\newcommand{\rsigma}{I\!\!R^{\sigma}}
\renewcommand{\rq}{I\!\!R^{q}}

\newcommand{\N}{{I\!\!N}}

\newcommand{\Z}{{Z\!\!\!Z}}
\newcommand{\eu}{{\cal E}^u}
\newcommand{\ebeta}{{\cal E}_\beta}

\newcommand{\thi}{\tilde{H}}
\newcommand{\cp}{\mathcal{P}}
\newcommand{\bnu}{\bar{\nu}}
\newcommand{\mboxdoms}{\mbox{dom} \, s}
\newcommand{\mboxemdoms}{\mbox{{\em dom}} \, s}
\newcounter{bean}
\newcommand{\benuma}{\setlength{\labelwidth}{.25in}
\begin{list}%
{(\alph{bean})}{\usecounter{bean}}}
\newcommand{\eenuma}{\end{list}}
\def\theequation{\thesection.\arabic{equation}}
\def\theequation{\arabic{section}.\arabic{equation}}
\def\thedefn{\arabic{section}.\arabic{defn}}
\newcommand{\beginsec}{\setcounter{equation}{0}}

\begin{document}

\title{Complete Analysis of Phase Transitions and Ensemble Equivalence
for the Curie-Weiss-Potts Model}
\author{Marius Costeniuc}
\affiliation{Department of Mathematics and Statistics, University of
Massachusetts, Amherst, MA, USA 01003}
\author{Richard S. Ellis}
\email{rsellis@math.umass.edu}
\affiliation{Department of Mathematics and Statistics, University of
Massachusetts, Amherst, MA, USA 01003}
\author{Hugo Touchette}
\email{htouchet@alum.mit.edu}
\affiliation{School of Mathematical Sciences,
Queen Mary, University of London, London, UK E1 4NS}

\begin{abstract}
Using the theory of large deviations, we analyze the phase transition structure
of the Curie-Weiss-Potts spin model, which is a mean-field
approximation to the Potts model.  This analysis is
carried out both for the canonical ensemble and the
microcanonical ensemble.  Besides giving explicit formulas for the
microcanonical entropy and for the equilibrium macrostates
with respect to the two ensembles, we analyze ensemble equivalence
and nonequivalence at the level of equilibrium macrostates, relating these to concavity
and support properties of the microcanonical entropy.
The Curie-Weiss-Potts model is the first statistical mechanical
model for which such a detailed and rigorous analysis has been carried out.
\end{abstract}


\maketitle

\section{Introduction}
The nearest-neighbor Potts model, introduced in \cite{Potts},
takes its place next to the Ising model as one
of the most versatile models in equilibrium statistical mechanics
\cite{Wu}.  Section I.C of \cite{Wu} presents a mean-field approximation to the 
Potts model, defined in terms of a mean interaction averaged over
all the sites in the model.  We refer to this approximation
as the Curie-Weiss-Potts model.  
Both the nearest-neighbor Potts model and the Curie-Weiss-Potts model 
are defined by sequences of probability distributions of $n$
spin random variables that may occupy one of $q$ different states $\theta^1,\ldots,
\theta^q$, where $q \geq 3$.  For $q=2$ the Potts model reduces to the Ising model while
the Curie-Weiss-Potts model reduces to the much simpler mean-field approximation to the
Ising model known as the Curie-Weiss model \cite{Ellis}.

Two ways
in which the Curie-Weiss-Potts model approximates the Potts model, and in 
fact gives rigorous bounds on quantities in the Potts model, are discussed
in \cite{KS} and \cite{PeaGri}.  Probabilistic limit theorems for the Curie-Weiss-Potts model 
are proved in \cite{EW1}, including the law of large numbers and its 
breakdown as well as various types of central limit theorems.  The model is also studied in \cite{EW2}, 
which focuses on a statistical estimation problem for two parameters defining the model.

In order to carry out the analysis of the model in \cite{EW1,EW2}, 
detailed information about the structure of the
set of canonical equilibrium macrostates is required, including the fact
that it exhibits a discontinuous phase transition as the inverse temperature
$\beta$ increases through a critical value $\beta_c$.  This information
plays a central role in the present paper, in which we 
use the theory of large deviations to study the equivalence and nonequivalence
of the sets of equilibrium macrostates for the microcanonical and canonical
ensembles.  An important consequence of the discontinuous
phase transition exhibited by the canonical ensemble in the Curie-Weiss-Potts
model is the implication that
the nearest-neighbor Potts model on $\Z^d$ also undergoes a discontinuous phase
transition whenever $d$ is sufficiently large \cite[Thm.\ 2.1]{BisCha}.

In \cite{EHT} the problem of the equivalence of the microcanonical
and canonical ensembles was completely solved
for a general class of statistical mechanical models including
short-range and long-range spin models and models of turbulence.  
This problem is
fundamental in statistical mechanics because it focuses on the
appropriate probabilistic description of statistical mechanical
systems.  While the theory developed in 
\cite{EHT} is complete, our understanding is greatly enhanced
by the insights obtained from studying specific models.  In this regard
the Curie-Weiss-Potts model is an excellent choice, lying at the boundary
of the set of models for which a complete analysis involving explicit formulas
is available.

For the Curie-Weiss-Potts model ensemble equivalence at the thermodynamic level is 
studied numerically in \cite[\S3--5]{IspCoh}.  This level of ensemble equivalence focuses
on whether the microcanonical entropy is concave on its domain; equivalently,
whether the microcanonical entropy and the canonical free energy, the 
basic thermodynamic functions in the two ensembles, can each be expressed
as the Legendre-Fenchel transform of the other \cite[pp.\ 1036--1037]{EHT}.  
Nonconcave anomalies in the microcanonical entropy 
partially correspond to regions of negative specific heat
and thus thermodynamic instability.  

The present paper significantly extends \cite[\S3--5]{IspCoh} by analyzing rigorously
ensemble equivalence at the thermodynamic level and by relating it to ensemble
equivalence at the level of equilibrium macrostates via the results in \cite{EHT}.
As prescribed by the theory of large deviations, the set $\eu$ of microcanonical equilibrium 
macrostates and the set $\ebeta$ of canonical equilibrium macrostates are
defined in (\ref{eqn:micromacro}) and (\ref{eqn:canonmacro}).
These macrostates are, respectively, the solutions
of a constrained minimization problem involving probability vectors on $\R^q$
and a related, unconstrained minimization problem.  The equilibrium macrostates for
the two ensembles are probability vectors describing equilibrium configurations
of the model in each ensemble in the thermodynamic limit $n \goto \infty$.  
For each $i = 1,2,\ldots,q$, the $i$th component of an equilibrium macrostate gives the asymptotic
relative frequency of spins taking the spin-value $\theta^i$.  

Defined via conditioning on $h_n$, the microcanonical ensemble expresses the conservation 
of physical quantities such as the energy.  Among other reasons, the mathematically more tractable 
canonical ensemble was introduced by Gibbs \cite{Gibbs} in the hope that
in the $n \goto \infty$ limit the two ensembles are equivalent; i.e., 
all asymptotic properties of the model obtained via the microcanonical ensemble
could be realized as asymptotic properties obtained via the canonical ensemble.  
Although most textbooks in statistical mechanics, including 
\cite{Bali,Gibbs,Huang,LanLif2,Reif,Sal}, claim that
the two ensembles always give the same predictions, in general this is not the case
\cite{TouEllTur}.   There are many examples of statistical mechanical models for which
nonequivalence of ensembles holds over a wide range of model
parameters and for which physically interesting microcanonical
equilibria are often omitted by the canonical ensemble.  Besides
the Curie-Weiss-Potts model, these models
include the mean-field Blume-Emery-Griffiths model \cite{BarMukRuf1,BarMukRuf2,EllTouTur},
the Hamiltonian mean-field model \cite{DauLatRapRufTor,LatRapTsa2},
the mean-field X-Y model \cite{DauHolRuf}, models of turbulence
\cite{CagLioMarPul1,EllHavTur3,EyiSpo,KieLeb,RobSom}, models of plasmas \cite{KieNeu2,SmiOne},
gravitational systems \cite{Gross1,Gross2,HerThi,LynBelWoo,Thi2}, and a model of the Lennard-Jones
gas \cite{BorTsa}. It is hoped that our detailed analysis of ensemble nonequivalence in the Curie-Weiss-Potts
model will contribute to an understanding of this fascinating and fundamental
phenomenon in a wide range of other settings.

In the present paper, after summarizing the large deviation analysis of the
Curie-Weiss-Potts model in Section 2,
we give explicit formulas for the elements of $\ebeta$ and the elements of $\eu$ in
Sections 3 and 4.  This analysis shows that $\ebeta$ exhibits
a discontinuous phase transition at a critical inverse temperature $\beta_c$
and that $\eu$ exhibits
a continuous phase transition at a critical mean energy $u_c$.
The implications of these different phase transitions concerning ensemble nonequivalence
are studied graphically in Section 5 and rigorously in Section 6, where we
exhibit a range of values of the mean energy for which the microcanonical
equilibrium macrostates are not realized canonically.  As described in
the main theorem in \cite{EHT} and summarized here in Theorem \ref{thm:equiv},
this range of values of the mean energy is precisely the set on
which the microcanonical entropy is not concave.  The analysis of
this bridge between ensemble nonequivalence at the thermodynamic level and
ensemble nonequivalence at the
level of equilibrium macrostates is one of the main contributions
of \cite{EHT} for general models and of the present paper for the Curie-Weiss-Potts
model.  In a sequel to the present paper \cite{CosEllTou}, we will extend our analysis of
the Curie-Weiss-Potts model to the so-called Gaussian ensemble
\cite{CH1,CH2,Heth,HethStump,JPV,StumpHeth} to show, among other
things, that for each value of the mean energy for which the
microcanonical and canonical ensembles are nonequivalent, we can find a
Gaussian ensemble that is fully equivalent with the microcanonical
ensemble \cite{CosEllTouTur}.

\section{Sets of Equilibrium Macrostates for the Two Ensembles}
\beginsec

Let $q \geq 3$ be a fixed integer and define
$\Lambda = \{\theta^1,\theta^2,\ldots,\theta^q\}$, where the
$\theta^i$ are any $q$ distinct vectors in $\R^q$.  
In the definition of the Curie-Weiss-Potts model, 
the precise values of these vectors is immaterial.
For each $n \in \N$ the model is defined by spin random variables
$\omega_1,\omega_2,\ldots,\omega_n$ that take values in $\Lambda$.
The canonical and microcanonical ensembles for the model are defined in terms of
probability measures on the configuration spaces $\Lambda^n$, which consist of
the microstates $\omega = (\omega_1,...,\omega_n)$.   
We also introduce the $n$-fold product measure $P_n$ 
on $\Lambda^n$ with identical one-dimensional marginals 
\[
\bar\rho = \frac{1}{q}\sum_{i=1}^q \delta_{\theta^i}.
\]
Thus for all $\omega \in \Lambda^n$, $P_n(\omega) = \frac{1}{q^n}$.
For $n \in \N$ and $\omega \in \Lambda^n$ 
 the Hamiltonian for the $q$-state Curie-Weiss-Potts model 
is defined by 
\[
H_n(\omega) = - \frac{1}{2n} \sum_{j,k=1}^n \delta(\omega_j,\omega_k),
\]
where $\delta(\omega_j,\omega_k)$ equals 1 if $\omega_j = \omega_k$ and
equals 0 otherwise.  The energy per particle is defined by
\[
\label{eqn:hn}
h_n(\omega) = \frac{1}{n} H_n(\omega).
\]

For inverse temperature $\beta \in \R$ and subsets $B$ of $\Lambda^n$
the canonical
ensemble is the probability measure $P_{n,\beta}$ defined by
\[
P_{n,\beta}\{B\} = \frac{1}{\sum_{\omega \in \Lambda^n} \exp[-n\beta h_n(\omega)]}
\cdot
\sum_{\omega \in B} \exp[-n\beta h_n(\omega)].
\]
For mean energy $u \in \R$ and $r > 0$ the microcanonical ensemble is the conditioned
probability measure $P_n^{u,r}$ defined by 
\[
P_n^{u,r}\{B\} = P_n\{B \, | \, h_n \in [u-r,u+r]\}.
\]
 The key to our analysis of the Curie-Weiss-Potts model is to express both the
canonical and the microcanonical ensembles in terms of the
empirical vector 
\[
L_n = L_n(\omega) = (L_{n,1}(\omega),L_{n,2}(\omega),\ldots,L_{n,q}(\omega)),
\
\]
the $i$th component of which is defined by
\[
\label{eqn:empmeasure}
L_{n,i}(\omega) = \frac{1}{n} \sum_{j=1}^n
\delta(\omega_j,\theta^i).
\]
This quantity equals the relative frequency with which 
$\omega_j, j \in \{1,\ldots,n\},$ equals $\theta^i$.
$L_n$ takes values in the set of probability vectors
\[
\mathcal{P} = \left\{\nu \in \R^q : \nu = (\nu_1,\nu_2,\ldots,\nu_q), \mbox{ each } \nu_i \geq 0,
\sum_{i=1}^q \nu_i = 1 \right\}.
\]
As we will see, each probability vector in $\cp$ represents a possible equilibrium
macrostate for the model.  

There is a one-to-one correspondence between $\cp$ and the set $\cp(\Lambda)$
of probability measures on $\Lambda$, $\nu \in \cp$ corresponding to the 
probability measure $\sum_{i=1}^q \nu_i \delta_{\theta^i}$.  The element
$\rho \in \cp$ corresponding to the one-dimensional marginal $\bar\rho$ of
the prior measures $P_n$ is the uniform vector having equal components $\frac{1}{q}$.

We denote by $\lan \cdot, \cdot \ran$ the inner product on $\R^q$.  Since
\[
\sum_{i=1}^q \sum_{j=1}^n \delta(\omega_j,\xi^i) \cdot \sum_{k=1}^n \delta(\omega_k,\xi^i) =
\sum_{j,k=1}^n \delta(\omega_j,\omega_k),
\]
it follows that the energy per particle can be rewritten as
\[
\label{eqn:empiricalmeasureenergy}
h_{n}(\omega) = - \frac{1}{2n^2} \sum_{j,k=1}^n \delta(\omega_j,\omega_k) =
-\ts\frac{1}{2} \lan L_n(\omega),L_n(\omega) \ran,
\]
i.e., 
\be 
\label{eqn:interactfcn}
h_n(\omega) = \thi(L_n(\omega)), \mbox{ where }
\thi(\nu) = -\ts\frac{1}{2} \lan \nu,\nu \ran  
\mbox{ for } \nu \in \mathcal{P}.
\ee
We call
$\thi$ the energy representation function.  

We appeal to the theory of large deviations to define the sets
of microcanonical equilibrium macrostates and canonical equilibrium macrostates.
Sanov's Theorem  states that with respect to the
product measures $P_n$, the empirical vectors
$L_n$ satisfy the large deviation principle (LDP) on $\cp$ with rate function
given by the relative entropy $R(\cdot|\rho)$ \cite[Thm.\ VIII.2.1]{Ellis}.  
For $\nu \in \cp$ this is defined by
\[
R(\nu|\rho) = \sum_{i=1}^q \nu_i \log (q\nu_i).
\]
We express this LDP by the formal notation $P_n\{L_n \in d\nu\} \approx
\exp[-n R(\nu|\rho)]$.  
The LDPs for $L_n$ with
respect to the two ensembles $P_{n,\beta}$ and $P_n^{u,r}$ 
in the thermodynamic limit $n \goto \infty$, $r \goto 0$ can be proved from the
LDP for the $P_n$-distributions of $L_n$ as in Theorems 2.4 and 3.2 in
\cite{EHT}, in which minor notational changes have to be made.
We express these LDPs by the formal notation
\be
\label{eqn:formal}
P_{n,\beta}\{L_n \in d\nu\} \approx \exp[-n I_\beta(\nu)]
\ \mbox{ and } \ 
P_n^{u,r}\{L_n \in d\nu\} \approx \exp[-n I^u(\nu)],
\ee
where for $\nu \in \cp$
\[
I_\beta(\nu) = R(\nu|\rho) - \ts \frac{\beta}{2}\lan \nu,\nu \ran - \mbox{const}
\]
and 
\[
I^u(\nu) = \left\{
\begin{array}{ll} R(\nu|\rho) - \mbox{const} & \ \mbox{ if } \: 
\ts -\frac{1}{2}\lan \nu,\nu \ran = u \\ \infty & \ \mbox{ otherwise. }
\end{array}
\right.
\]
The constants appearing in the definitions of $I_\beta$ and $I^u$ have
the properties that $\inf_{\nu \in \cp}I_\beta(\nu) = 0$
and $\inf_{\nu \in \cp}I^u(\nu) = 0$.  Thus
$I_\beta$ and $I^u$ map ${\cal P}$ into $[0,\infty)$.

As the formulas in (\ref{eqn:formal}) suggest, if $I_\beta(\nu) > 0$ or $I^u(\nu) > 0$, then
$\nu$ has an exponentially small probability of being observed in the corresponding
ensemble in the thermodynamic limit.  Hence it makes sense to define the
corresponding sets of equilibrium macrostates to be
\[
\ebeta = \{\nu \in \cp : I_\beta(\nu) = 0\} \ \mbox{ and } \ 
\eu = \{\nu \in \cp : I^u(\nu) = 0\}.
\]
A rigorous justification for this is given in \cite[Thm.\ 2.4(d)]{EHT}.
Using the formulas for $I_\beta$ and $I^u$, we see that
\be
\label{eqn:canonmacro}
\ebeta =  \left\{\nu \in
\mathcal{P} : \nu \mbox{ minimizes } R(\nu |\rho) - \ts \frac{\beta}{2}\lan
\nu,\nu \ran\right\}
\ee
and
\be
\label{eqn:micromacro}
\eu = 
\left\{ \nu \in \mathcal{P}
: \nu \mbox{ minimizes } R(\nu | \rho) \mbox{ subject to } 
- \ts \frac{1}{2} \lan \nu,\nu \ran = u \right\}.
\ee
Each element $\nu$ in $\ebeta$ and $\eu$ 
describes an equilibrium configuration of the model in the corresponding
ensemble in the thermodynamic limit.
The $i$th component $\nu_i$ gives the asymptotic relative
frequency of spins taking the value $\theta^i$.

The question of equivalence of ensembles at the level of equilibrium macrostates
focuses on the relationships between $\eu$, defined in terms of the constrained minimization
problem in (\ref{eqn:micromacro}), and $\ebeta$, defined in terms of the related,
unconstrained minimization problem in (\ref{eqn:canonmacro}).
We will focus on this question in Sections 5 and 6 after we determine the structures
of $\ebeta$ and $\eu$ in the next two sections.

\section{Form of $\ebeta$ and Its Discontinuous Phase Transition}
\beginsec

In this section we derive the form of the set $\ebeta$ of canonical equilibrium
macrostates for all $\beta \in \R$.  This form is given in 
Theorem \ref{thm:eb}, which shows that with respect
to the canonical ensemble the Curie-Weiss-Potts model undergoes
a discontinuous phase transition at the critical inverse temperature 
\be
\label{eqn:betac}
\beta_c = \frac{2(q-1)}{q-2} \log(q-1).
\ee
In order to describe the form of $\ebeta$, 
we introduce the function $\psi$ that maps $[0,1]$ into $\cp$ and is defined by
\be
\label{eqn:psi}
\psi(w) =  \left(\frac{1+(q-1)w}{q}, \frac{1-w}{q}, \ldots,
\frac{1-w}{q}\right);
\ee
the last $q-1$ components all equal $\frac{1-w}{q}$.
Recalling that $\rho$ is the uniform vector in $\cp$ having equal components
$\frac{1}{q}$, we see that $\rho = \psi(0)$.  

\begin{theorem}\per
\label{thm:eb}  
For $\beta > 0$ let $w(\beta)$ be the largest solution of the equation
\be
\label{eqn:wbeta}
w = \frac{1-e^{-\beta w}}{1+(q-1)e^{-\beta w}}.
\ee
The following conclusions hold.

{\em (a)} The quantity $w(\beta)$ is well defined and lies in $[0,1]$.  
It is positive, strictly increasing, and differentiable
for $\beta \in (\beta_c, \infty)$ and
satisfies $w(\beta_c)= \frac{q-2}{q-1}$ and $\lim_{\beta
\goto \infty} w(\beta)=1$.  

{\em (b)}  For $\beta \geq \beta_c,$ define
$\nu^1(\beta)=\psi(w(\beta))$ and let $\nu^i(\beta)$, $i=2, \ldots, q$,
denote the points in $\R^q$ obtained by interchanging the first and $i$th
components of $\nu^1(\beta)$.  Then the set $\ebeta$ defined in
{\em (\ref{eqn:canonmacro})} has the form
\be
\label{eqn:ebetaexplicit}
{\cal{E}}_\beta = \left\{
\begin{array}{ll}
           \{\rho\}
 & \ \mbox{ for } \: \beta < \beta_c \, , 
\\
           \{\rho, \nu^1(\beta_c), \nu^2(\beta_c), \ldots,
\nu^q(\beta_c)\}
 & \ \mbox{ for }  \: \beta=\beta_c \, ,  \rule{0in}{.2in}
\\
           \{\nu^1(\beta), \nu^2(\beta), \ldots, \nu^q(\beta)\}
 & \ \mbox{ for }  \: \beta > \beta_c \, .  \rule{0in}{.2in}
\end{array}
\right.
\ee
For $\beta \geq \beta_c,$ the vectors in ${\cal{E}}_\beta$ are all
distinct and each $\nu^i(\beta)$ is continuous. The vector $\nu^1(\beta_c)$ is given by
\be
\label{eqn:nu1betac}
\nu^1(\beta_c) = \psi(w(\beta_c)) =  \psi\!\left(\ts \frac{q-2}{q-1}\right) =
\left(\ts 1 - \frac{1}{q}, \frac{1}{q(q-1)}, \ldots,
\frac{1}{q(q-1)}\right);
\ee
the last $q-1$ components all equal $\frac{1}{q(q-1)}$.  
\end{theorem}

The form of $\ebeta$ for $\beta > 0$
is proved in Appendix B from a new convex-duality theorem proved in Appendix A
and from the complicated calculation of the global minimum points of a related function 
given in Theorem 2.1 in \cite{EW1}.  The form of $\ebeta$ for $\beta \leq 0$ 
is also determined in Appendix B.

For $\beta > 0$ the form of $\ebeta$ reflects a competition
between disorder, as represented by the relative entropy $R(\nu|\rho)$, and 
order, as represented by the energy representation
function $-\frac{1}{2}\lan \nu,\nu \ran$.  For small $\beta > 0$,
$R(\nu|\rho)$ predominates.  Since $R(\nu|\rho)$ attains its minimum of 0 at the unique vector
$\rho$, we expect that for small $\beta$, $\ebeta$ should contain 
a single vector.  On the other hand, for large $\beta > 0$, $-\frac{1}{2}\lan \nu,\nu \ran$
predominates.  This function attains its minimum
at $\nu^1 = (1,0,\ldots,0)$ and at the vectors $\nu^i$, $i=1,\ldots,q$, obtained
by interchanging the first and $i$th components of $\nu^1$.  Hence we expect
that for large $\beta$, $\ebeta$ should contain $q$ distinct
vectors $\nu^i(\beta)$ having the property that $\nu^i(\beta) 
\goto \nu^i$ as $\beta \goto \infty$.  
The major surprise of the theorem is that for $\beta = \beta_c$,
$\ebeta$ consists of the $q+1$ distinct vectors $\rho$
and $\nu^i(\beta_c)$ for $i=1,2,\ldots,q$.  

The discontinuous bifurcation in 
the composition of $\ebeta$ from 1 vector for $\beta < \beta_c$
to $q+1$ vectors for $\beta = \beta_c$ to $q$ vectors for
$\beta > \beta_c$ corresponds to a discontinuous
phase transition exhibited by the canonical ensemble.
In Figure 2 in Section 5 this
phase transition is shown together with the continuous
phase transition exhibited by the microcanonical ensemble.   The
latter phase transition and the form of the set of microcanonical
equilibrium macrostates are the focus of the next section.

\section{Form of $\eu$ and Its Continuous Phase Transition}
\beginsec

We now turn to the form of the set $\eu$ for all
$u \in [-\frac{1}{2},-\frac{1}{2q}]$, which is the set of
$u$ for which $\eu$ is nonempty.
In the specific case $q=3$ part (c) of Theorem \ref{thm:eu} gives the form of $\eu$,
the calculation of which is much simpler than the calculation of the form of $\ebeta$.
The proof is based on the method of Lagrange multipliers,
which also works for general $q \geq 4$ provided the next conjecture
on the form of the elements in $\eu$ is valid.
The validity of this conjecture has been confirmed numerically
for all $q \in \{4,5,\ldots,10^4\}$
and all $u \in (-\frac{1}{2},-\frac{1}{2q})$ of the
form $u = -\frac{1}{2} + 0.02 k$, where $k$ is a positive integer.

\begin{conj}\per
\label{conj:eu}
For any $q \geq 4$ and all $u \in (-\frac{1}{2},-\frac{1}{2q})$, there exists $a \not = b \in (0,1)$
such that modulo permutations, any $\nu \in \eu$ has the form
$(a,b,\ldots,b)$; the last $q-1$ components of which all equal $b$.
\end{conj}

Parts (a) and (b) of Theorem \ref{thm:eu} are proved for general $q \geq 3$.
Part (c) shows that modulo permutations, for
$q=3$, $\nu \in \eu$ has the form $(a(u), a(u), b(u))$ and determines the precise
formulas for $a(u)$ and $b(u)$.  As specified in part (d), for $q \geq 4$ we can also determine
the precise formula for $\nu \in \eu$ provided Conjecture \ref{conj:eu} is valid.

Theorem \ref{thm:eu} shows that with respect to the microcanonical ensemble
the Curie-Weiss-Potts model undergoes a continuous phase
transition as $u$ decreases from the critical mean-energy value $u_c = -\frac{1}{2q}$.
This contrast with the discontinuous phase transition 
exhibited by the canonical ensemble is closely related to the
nonequivalence of the microcanonical and canonical ensembles for a range of $u$.
Ensemble equivalence and nonequivalence will be explored in the next section,
where we will see that it is reflected by support and concavity properties
of the microcanonical entropy.  An explicit formula for the microcanonical entropy 
is given in Theorem \ref{thm:su}.

\begin{thm}\per
\label{thm:eu}
For $u \in \R$ we define $\eu$ by {\em (\ref{eqn:micromacro})}.  The following conclusions hold.

{\em (a)} For any $q \geq 3$, $\eu$ is nonempty if and only if $u \in [-\frac{1}{2},-\frac{1}{2q}]$.  
This interval coincides with the range of the energy representation
function $\thi(\nu) = -\frac{1}{2}\lan \nu,\nu \ran$ on $\cp$.

{\em (b)} For any $q \geq 3$, ${\cal E}^{-\frac{1}{2q}} = \{\rho\} =
\{(\frac{1}{q},\frac{1}{q},\ldots \frac{1}{q})\}$ and
\[
{\cal E}^{-\frac{1}{2}} = \{(1,0,\ldots,0), (0,1,\ldots,0), \ldots,
(0,0,\ldots,1)\}.
\]

{\em (c)} Let $q=3$.  
For $u \in (-\frac{1}{2},-\frac{1}{2q})$,
${\cal{E}}^u$ consists of the $3$ distinct vectors $\{ \nu^1(u), \nu^2(u), \nu^3(u) \}$, where 
$\nu^1(u) =(a(u), b(u), b(u))$,
\be
\label{eqn:aubu}
a(u)=   \frac{1+\sqrt{2(-6u-1)} }{3} \ \ \mbox{ and } \ \
b(u)=  \frac{2-\sqrt{2(-6u-1)} }{6}.
\ee
The vectors $\nu^2(u)$ and $\nu^3(u)$ denote the points in $\R^3$ obtained by
interchanging the first and the $i$th components of $\nu^1(u)$.

{\em (d)}  Let $q \geq 4$ and assume that Conjecture {\em \ref{conj:eu}}
is valid.  Then for $u \in (-\frac{1}{2}, -\frac{1}{2q})$,
${\cal{E}}^u$ consists of the $q$ distinct vectors $\{ \nu^1(u), \ldots, \nu^q(u) \}$, where
$\nu^1(u) =(a(u), b(u), \ldots, b(u))$,
\[
a(u) =  \frac{1+\sqrt{(q-1)(-2qu-1)} }{q} \ \ \mbox{ and } \ \
b(u)=  \frac{q-1-\sqrt{(q-1)(-2qu-1)} }{(q-1)q}.
\]
The last $q-1$ components of $\nu^1(u)$ all equal $b(u)$, 
and the vectors $\nu^i(u), i=2, \ldots, q$, denote the points in $\R^q$ obtained by
interchanging the first and the $i$th components of $\nu^1(u)$.  
\end{thm}

We return to part (b) of Theorem \ref{thm:eu} in order to
discuss the nature of the phase transition exhibited by the microcanonical ensemble.
The functions $a(u)$ and $b(u)$ given in (\ref{eqn:aubu})
are both continuous for $u \in [-\frac{1}{2},-\frac{1}{2q}]$
and satisfy
\[
\lim_{u \goto -\frac{1}{2q}} a(u) = \lim_{u \goto -\frac{1}{2q}} b(u)
= \ts \frac{1}{q} = a(-\frac{1}{2q}) = b(-\frac{1}{2q}).
\]
Therefore, for $i = 1,\ldots,q$, $\lim_{u \goto -\frac{1}{2q}} \nu^i(u) = \rho$.
It follows that the microcanonical ensemble exhibits a continuous phase
transition as $u$ decreases from $u_c = -\frac{1}{2q}$, 
the unique equilibrium macrostate $\rho$ for $u = u_c$
bifurcating continuously into the $q$ distinct macrostates $\nu^{(i)}(u)$ as $u$
decreases from its maximum value.  This is rigorously true for $q=3$.  Provided
Conjecture \ref{conj:eu} is true, it is also true for $q \geq 4$, as one 
easily checks using part (d) of Theorem \ref{thm:eu}.
  
Before proving Theorem \ref{thm:eu}, we introduce the microcanonical entropy
\be
\label{eqn:microentropy}
s(u) = -\inf \!\left\{ R(\nu | \rho) : \nu \in \mathcal{P}, -\ts\frac{1}{2}
\lan \nu,\nu \ran = u \right\}.
\ee 
As we will see in the next section, this function plays a crucial role in
the analysis of ensemble equivalence and nonequivalence for the Curie-Weiss-Potts model.
Since $0 \leq R(\nu|\rho)$ for all $\nu \in \cp$, $s(u) \in [-\infty,0]$ for all $u$,
and since $R(\nu|\rho) > R(\rho|\rho) = 0$ for all $\nu \not = \rho$, $s$ attains
its maximum of 0 at the unique value $-\frac{1}{2q} = -\frac{1}{2}\lan \rho,\rho \ran$.

The domain of $s$ is defined by $\mbox{dom} \, s = \{u \in \R : s(u) > -\infty\}$.
Since $R(\nu|\rho) < \infty$ for all $\nu \in \cp$, $\mbox{dom} \, s$ equals
the range of $\thi(\nu) = -\frac{1}{2}\lan \nu,\nu \ran$ on $\cp$, which is
the interval $[-\frac{1}{2},-\frac{1}{2q}]$ [Thm.\ \ref{thm:eu}(a)].
As we have seen, $s(-\frac{1}{2q}) = 0$.  
For $u \in (-\frac{1}{2},-\frac{1}{2q})$, according to parts (c)--(d) of Theorem \ref{thm:eu}
$\eu$ consists of the unique vector $\nu^{(1)}(u)$ modulo permutations.
  Since for $i = 2,3,\ldots,q$, 
$R(\nu^{(i)}(u)|\rho) = R(\nu^{(1)}(u)|\rho)$, we conclude that
\[
s(u) = - R(\nu^{(1)}(u)|\rho) = -a(u) \log(q \, a(u)) - (q-1) b(u) \log(q \, b(u)).
\]
Finally, for $u = -\frac{1}{2}$,
modulo permutations $\eu$ consists of the unique vector $(1,0,\ldots,0)$ [see
(\ref{eqn:qpoints})], and so $s(-\frac{1}{2}) = -R((1,0,\ldots,0)|\rho) = -\log q$.
The resulting formulas for $s(u)$ are recorded in the next theorem, where
we distinguish between $q = 3$ and $q \geq 4$.

\begin{thm}\per
\label{thm:su}
We define the microcanonical entropy $s(u)$ in {\em (\ref{eqn:microentropy})}.
The following conclusions hold.

{\em (a)} $\mboxemdoms = [-\frac{1}{2},-\frac{1}{2q}]$; for any $u \in
\mboxemdoms$, $u \not = -\frac{1}{2q}$,
$s(u) < s(-\frac{1}{2q}) = 0$; and $s(-\frac{1}{2}) = - \log q$.

{\em (b)} Let $q=3$.  Then for $u \in (-\frac{1}{2}, -\frac{1}{2q}) = (-\frac{1}{2}, -\frac{1}{6})$ 
\bea
\label{eqn:explicitsu3}
s(u) &=&
 - \ \frac{1+\sqrt{2(-6u-1)} }{3} \: \log\!\left(1+\sqrt{2(-6u-1)}\right) \\
 && \hspace{.1in} - \ \frac{2-\sqrt{2(-6u-1)}}{3} \: 
\log \!\left(\frac{2-\sqrt{2(-6u-1)}}{2}\right).
\nonumber 
\eea

{\em (c)}  Let $q \geq 4$ and assume that Conjecture {\em \ref{conj:eu}}
is valid.  Then for $u \in (-\frac{1}{2}, -\frac{1}{2q})$
\bea
\label{eqn:explicitsuq}
s(u) & = & 
 - \ \frac{1+\sqrt{(q-1)(-2qu-1)} }{q} \: \log\!
\left(1+\sqrt{(q-1)(-2qu-1)}\right) \\
  && \hspace{.1in} - \  \frac{q-1-\sqrt{(q-1)(-2qu-1)} }{q} \: \log\!
\left(\frac{q-1-\sqrt{(q-1)(-2qu-1)} }{q-1}\right). \nonumber
\eea
\end{thm}

We now turn to the proof of Theorem \ref{thm:eu}, which gives the form of $\eu$.
We start by proving part (a).  
The set $\rule{0in}{.15in}\eu$ of microcanonical equilibrium macrostates consists of all
$\nu \in \cp$ that minimize the relative entropy $R(\nu|\rho)$ subject to 
the constraint that
\[
\thi(\nu) = -\ts \frac{1}{2} \lan \nu,\nu \ran = u.
\]
Let $u = -\frac{1}{2}r^2$.
Since $\cp$ consists of all nonnegative vectors in $\r^q$ satisfying
$\nu_1 + \cdots + \nu_q = 1$, the constraint set in the minimization problem
defining $\eu$ is given by
\be
\label{eqn:cu}
C(u) = C(\ts -\frac{1}{2}r^2) = \ds
\left\{\nu \in \R^q : \nu_1 \geq 0, \ldots, \nu_q \geq 0, \: \sum_{j=1}^q \nu_j = 1, \:
\sum_{j=1}^q \nu_j^2 = r^2\right\}.
\ee
Geometrically, $C(-\frac{1}{2} r^2)$ is the intersection of the 
nonnegative orthant of $\r^q$, the hyperplane
consisting of $\nu \in \rsigma$ that satisfy
$\nu_1 + \cdots + \nu_q = 1$, and the hypersphere in $\rq$ with center 0 and radius $r$.
Clearly, $C(u) \not = \emptyset$ if and only if $u$ lies in the range of the 
energy representation function
$\thi(\nu) = -\frac{1}{2}\lan \nu,\nu \ran$ on $\cp$.
Because $0 \leq R(\nu|\rho) < \infty$ for all $\nu \in C(u)$, the
range of $\thi$ on ${\cal P}$ 
also equals the set of $u$ for which $\eu \not = \emptyset$.

The geometric description of $C(u)$ makes it straightforward to determine
those values of $u$ for which this constraint set is nonempty.  
The smallest value of $r$ for which $C(-\frac{1}{2} r^2) \not =
\emptyset$ is
obtained when the hypersphere of radius $r$ is tangent to the hyperplane,
the point of tangency being $\rho = (\frac{1}{q}, \frac{1}{q}, \ldots, \frac{1}{q})$,
the closest probability vector to the origin.  
The hypersphere and the hyperplane are tangent when $r = \frac{1}{\sqrt{q}}$, which coincides
with the distance from
the center of the hypersphere to the hyperplane.  It follows that
the largest value of $u$ for which $C(u) \not = \emptyset$, and
thus $\eu \not = \emptyset$, is $u = -\frac{1}{2}r^2 = -\frac{1}{2q}$.
In this case
\be
\label{eqn:onepoint}
C(\ts-\frac{1}{2q}) = \{\rho\} = \{(\frac{1}{q}, \frac{1}{q}, \ldots, \frac{1}{q})\}
= {\cal E}^{-\frac{1}{2q}}.
\ee

For all sufficiently large $r$,
$C(-\frac{1}{2} r^2)$ is empty because the 
hypersphere of radius $r$ has empty intersection with the intersection
of the hyperplane and the nonnegative orthant of $\r^q$.
The largest value for $r$ for which this does not occur
is found 
by subtracting the two equations defining the hyperplane and the hypersphere.
Since each $\nu_i \in [0,1]$, it follows that
\[
0 \leq \sum_{i=1}^q \nu_i(1-\nu_i) = 1 - r^2,
\]
and this in turn implies that $r^2 \leq 1$.  Thus $r = 1$ is the
largest value for $r$ for which $C(-\frac{1}{2} r^2)
\not = \emptyset$.  We conclude that the smallest 
value of $u$ for which $C(u) \not = \emptyset$, and thus 
$\eu \not = \emptyset$, is $u = -\frac{1}{2}r^2 = -\frac{1}{2}$.  
The set ${\cal E}^{-\frac{1}{2}}$ consists of the points
at which the hyperplane intersects each of the positive
coordinate axes; i.e.,
\be
\label{eqn:qpoints}
{\cal E}^{-\frac{1}{2}} = \{(1,0,\ldots,0), (0,1,\ldots,0), \ldots,
(0,0,\ldots,1)\}.
\ee
This completes the proof of part (a) of Theorem \ref{thm:eu}.

We now determine the form $\eu$ as specified in parts (b)--(d) of Theorem \ref{thm:eu}.
Part (b) considers any $q \geq 3$ and the values $u = -\frac{1}{2q}$ and $u = -\frac{1}{2}$,
part (c) $q = 3$ and $u \in (-\frac{1}{2},-\frac{1}{2q})$, and 
part (d) $q \geq 4$ and $u \in (-\frac{1}{2},-\frac{1}{2q})$.
Part (b) has already been proved; for $u = -\frac{1}{2q}$ and $u = -\frac{1}{2}$, the sets 
$\eu$ are given in (\ref{eqn:onepoint}) and (\ref{eqn:qpoints}).

We now consider $q \geq 3$ and $u \in (-\frac{1}{2},-\frac{1}{2q})$.
For $\nu \in {\cal P}$ define
\[
K(\nu) = \sum_{j=1}^q \nu_j \ \mbox{ and } \ \thi(\nu) = - \ts \frac{1}{2} \ds \sum_{j=1}^q \nu_j^2.
\]
By definition $\nu = (\nu_1, \ldots, \nu_q) \in {\cal{E}}^u$ if and only if
$\nu$ minimizes $R(\nu|\rho) = \sum_{j=1}^q \nu_j \log(q \nu_j)$ subject to the constraints
$K(\nu) = 1$, $\tilde{H}(\nu) = u$, and $\nu_1 \geq 0, \ldots, \nu_q \geq 0$.
For $u \in (-\frac{1}{2},-\frac{1}{2q})$ we divide into two parts
the calculation of the form of $\nu \in \eu$.  First we use
Lagrange multipliers to solve the constrained minimization problem 
when $\nu_1 > 0, \ldots, \nu_1 > 0$.  Then we
argue that the vectors $\nu$ found via Lagrange multipliers
solve the original constrained minimization problem when 
$\nu_1 \geq 0, \ldots, \nu_q \geq 0$.

We introduce Lagrange multipliers $\gamma$ and $\lambda$.
Any critical point of $R(\nu|\rho)$ subject to the constraints
$K(\nu) = 1$, $\thi(\nu) = u$, and $\nu_1 > 0, \nu_2 > 0, \ldots, \nu_q > 0$
satisfies
\[
\left\{
\begin{array}{ll}
           \nabla R(\nu|\rho) = \gamma \nabla K(\nu) + \lambda \nabla \tilde{H}(\nu) 
\\
           K(\nu)=1
\\
           \tilde{H}(\nu)=u
\\
           \nu_j > 0 \ \mbox{ for } j=1, 2, \ldots, q.
\end{array}
\right.
\]
This system of equations is equivalent to
\be
\label{eqn:fourlines}
\left\{
\begin{array}{ll}
            1+ \log(q \nu_j) = \gamma + \lambda \nu_j \ \mbox{ for }
j = 1, 2, \ldots ,q 
\\
               \sum_{j=1}^q \nu_j = 1
\\
            -\frac{1}{2} \sum_{j=1}^q \nu_{j}^{2} = u \rule{0in}{.175in}
\\
\nu_j > 0 \ \mbox{ for } j =1, 2, \ldots, q.
\end{array}
\right.
\ee
By properties of the logarithm, the first equation can have at most two 
solutions.  Hence modulo permutations, there exists $n \in \{0,1,\ldots,q\}$
and distinct numbers $a, b \in (0,1)$ such
that the first $n$ components of any critical point $\nu$ all equal $a$ and the
last $q-n$ components of $\nu$ all equal $b$.
The second and third equations in (\ref{eqn:fourlines}) take the form
\be
\label{eqn:ab}
na + (q-n)b = 1 \ \mbox{ and } \ na^{2} + (q-n)b^{2} = -2u.
\ee

If $n = 0$, then $b = \frac{1}{q}$, while if $n = q$, then $a = \frac{1}{q}$.  
Both cases correspond to $\nu = (\frac{1}{q},\ldots,\frac{1}{q}) = \rho$ and
$u = -\frac{1}{2q}$, which does not lie in the open interval 
$(-\frac{1}{2},-\frac{1}{2q})$ currently under consideration.

We now consider $1 \leq n \leq q-1$.  In this case the two solutions of (\ref{eqn:ab}) are
\be
\label{eqn:a1nb1n}
a_1(n) =  \frac{n-\sqrt{n(q-n)(-2qu-1)} }{nq}, \ \ b_1(n) =
 \frac{q-n+\sqrt{n(q-n)(-2qu-1)} }{(q-n)q},
\ee
and
\be
\label{eqn:a2nb2n}
a_2(n) = \frac{n+\sqrt{n(q-n)(-2qu-1)} }{nq}, \ \ b_2(n) =
\frac{q-n-\sqrt{n(q-n)(-2qu-1)} }{(q-n)q}.
\ee
Since $u \in (-\frac{1}{2},-\frac{1}{2q})$, these quantities are all well defined 
and $a_j(n) \not = b_j(n)$ provided $u < -\frac{1}{2q}$.

We now specialize to $q=3$, the case considered in part (c) of Theorem \ref{thm:eu}.  
When $q = 3$, the interval
$(-\frac{1}{2},-\frac{1}{2q})$ equals $(-\frac{1}{2},-\frac{1}{6})$,
and we have $n \in \{1,2\}$.  Equations (\ref{eqn:a1nb1n}) and (\ref{eqn:a2nb2n}) take the form
\[
a_1(n) = \frac{n-\sqrt{n(3-n)(-6u-1)}}{3n}, \ \ \
b_1(n) = \frac{3-n+\sqrt{n(3-n)(-6u-1)}}{3(3-n)}
\]
and
\[
a_2(n) = \frac{n+\sqrt{n(3-n)(-6u-1)}}{3n}, \ \ \
b_2(n) = \frac{3-n-\sqrt{n(3-n)(-6u-1)}}{3(3-n)}.
\]
Any critical point $\nu$ either has $n$ components equal to $a_1(n)$ and
$q-n$ components equal to $b_1(n)$ or has $n$ components equal to $a_2(n)$ and 
$q-n$ components equal to $b_2(n)$.

Modulo permutations, the value $n=1$ corresponds to
\[
\nu = (a_1(1), b_1(1), b_1(1)) \ \ \mbox{ or } \ \ \nu = (a_2(1), b_2(1), b_2(1)),
\]
and the value $n = 2$ corresponds to
\[
\nu = (a_1(2), a_1(2), b_1(2)) \ \ \mbox{ or } \ \ \nu = (a_2(2), a_2(2), b_2(2)).
\]
For $j \in \{1,2\}$, one easily checks that 
\[
a_j(1) = b_{3-j}(2) \ \ \mbox{ and } \ \ a_j(2) = b_{3-j}(1).
\]
Thus, modulo permutation $(a_1(1), b_1(1), b_1(1))=(a_2(2), a_2(2), b_2(2))$ and 
$\ (a_2(1), b_2(1), b_2(1)) = (a_1(2), a_1(2),b_1(2))$, and so
modulo permutations, $n =1$ and $n=2$ yield the same points.
This shows that it suffices to consider only the case $n=1$.  
Since for all $u \in (-\frac{1}{2},-\frac{1}{6})$
\[
R(\,(a_2(1), b_2(1), b_2(1))\,|\,\rho) < R(\,(a_1(1), b_1(1), b_1(1)))\,|\,\rho),
\]
we conclude that modulo permutation $\nu = (a_2(1), b_2(1), b_2(1))$
is the unique minimizer of $R(\nu|\rho)$ subject to the constraints
$K(\nu) = 1$, $\thi(\nu) = u$, and $\nu_1 > 0, \nu_2 > 0, \nu_3 > 0$.

We now prove for $q = 3$ that the minimizers found via Lagrange multipliers
when $\nu_1 > 0, \nu_2 > 0, \nu_3 > 0$
also minimize $R(\nu|\rho)$ subject to the constraints
$K(\nu) = 1$, $\thi(\nu) = u$, and $\nu_1 \geq 0, \nu_2 \geq 0, \nu_3 \geq 0$.
If $\nu=(\nu_1, \nu_2, \nu_3)$ satisfies the 
constraints and has two components equal to zero, then modulo permutations
$\nu = (1,0,0)$ and $\thi(\nu) = u = -\frac{1}{2}$, which does not 
lie in the open interval $(-\frac{1}{2}, -\frac{1}{6})$
currently under consideration.  Thus we only have to consider the case 
where $\nu$ has one component equal to zero; i.e, $\nu=(0, a_0, b_0)$ with $a_0 \geq b_0$.  In this case
the second and third equations in (\ref{eqn:fourlines}) have the solution
\[
a_0 =  \frac{1+\sqrt{-4u-1}}{2}, \ \ b_0 =  \frac{1-\sqrt{-4u-1}}{2}.
\]
We now claim that modulo permutations the unique minimizer of
$R(\nu|\rho)$ subject to the constraints
$K(\nu) = 1$, $\thi(\nu) = u$, and $\nu_1 \geq 0, \nu_2 \geq 0, \nu_3 \geq 0$
has the form $(a_2(1), b_2(1), b_2(1))$ found in the preceding paragraph.
The claim follows from the calculation
\[
R(\,(a_2(1), b_2(1), b_2(1))\,|\, \rho) < R(\,(0, a_0, b_0)\,|\, \rho),
\]
which is valid for all $u \in (-\frac{1}{2},-\frac{1}{6})$.
This completes the proof of part (c) of Theorem \ref{thm:eu}, which gives the form
of $\nu \in \eu$ for $q = 3$ and $u \in (-\frac{1}{2},-\frac{1}{6})$.

We now turn to part (d) of Theorem \ref{thm:eu}, which gives the form of $\eu$\
for $q \geq 4$ and $u \in (-\frac{1}{2},-\frac{1}{2q})$.  If, as in the case $q=3$,
we knew that modulo permutations, the minimizers have the form $(a,b,\ldots,b)$
as specified in Conjecture \ref{conj:eu}, then as in the case $q = 3$
we would be able to derive explicit formulas for these minimizers.  
If Conjecture \ref{conj:eu} is true, then it is easily verified that modulo
permutations, $\eu$ consists of the 
unique point $\nu = (a_2(1),b_2(1),\ldots,b_2(1))$, where $a_2(1)$ and $b_2(1)$ 
are defined in (\ref{eqn:a2nb2n}) for $u \in (-\frac{1}{2},-\frac{1}{2q})$.  
This gives part (d) of Theorem \ref{thm:eu}.  The proof of the theorem is complete.

At the end of Section 6 we will see that there exists an explicit value of $u_0 \in (-\frac{1}{2},
-\frac{1}{2q})$ such that Conjecture \ref{conj:eu} is valid for any $q \geq 4$
and all $u \in (-\frac{1}{2},u_0]$.  Hence for these values of $u$ the form 
of $\nu \in \eu$ given in part (d) of Theorem \ref{thm:eu} and the formula
for $s(u)$ given in part (c) of Theorem \ref{thm:su} are both rigorously true.

\section{Equivalence and Nonequivalence of Ensembles}
\beginsec

As we saw in Section 3, the set $\ebeta$ of canonical equilibrium macrostates
undergoes a discontinuous phase transition as $\beta$ increases through
$\beta_c = \frac{2(q-1)}{q-2}\log(q-1)$, the unique
macrostate $\rho$ bifurcating discontinuously into the $q$ distinct macrostates $\nu^{(i)}(\beta)$.
By contrast, as we saw in Section 4, the set $\eu$ of microcanonical equilibrium macrostates
undergoes a continuous phase transition as $u$ decreases from
$u_c = -\frac{1}{2q}$, the unique
macrostate $\rho$ bifurcating continuously into the $q$ distinct macrostates $\nu^{(i)}(u)$.
The different continuity properties of these phase transitions shows already that the
canonical and microcanonical ensembles are nonequivalent. In this section
we study this nonequivalence in detail and
relate the equivalence and nonequivalence of these two sets of equilibrium
macrostates to concavity and support properties of the microcanonical entropy $s$
defined in (\ref{eqn:microentropy}).  This is done with the help of Figure 2, which is based on
the form of $s$ in Figure 1 and
on the results on ensemble equivalence and nonequivalence
in Theorem \ref{thm:equiv}.  In Figures 3 and 4 at the end of
the section we give, for $q = 3$, a beautiful geometric representation of $\ebeta$ and $\eu$ that
also shows the ensemble nonequivalence for a range of $u$.

We start by stating in Theorem \ref{thm:equiv}
results on ensemble equivalence and nonequivalence for the Curie-Weiss-Potts model.
Analogous results are derived in Theorems 4.4, 4.6, and 4.8 in \cite{EHT}
for a wide range of statistical mechanical models, of which the Curie-Weiss-Potts model
is a special case.
For $u \in \mbox{dom} \, s$ the possible
relationships between $\eu$ and $\ebeta$, 
given in part (a) of Theorem \ref{thm:equiv}, are that either the ensembles
are fully equivalent, partially equivalent, or nonequivalent.  Since
by part (b) canonical
equilibrium macrostates are always realized microcanonically
and since, by part (a)(iii), microcanonical
equilibrium macrostates are in general not realized canonically, it follows
that the microcanonical ensemble is the richer of the two ensembles. 

\begin{thm}\per
\label{thm:equiv} 
We define $s$ by {\em (\ref{eqn:microentropy})} and $\ebeta$ and $\eu$
by {\em (\ref{eqn:canonmacro})} and {\em (\ref{eqn:micromacro})}.
The following conclusions hold.

{\em (a)}  For fixed $u \in \mbox{{\em dom}} \, s$ one of the following
three possibilities occurs.

\hspace{.25in} {\em (i)} \mbox{{\em {\bf Full equivalence.}}} $\:$ There exists
$\beta \in \R$ such that ${\cal E}^u = {\cal E}_\beta$.  This is the case if and
only if $s$ has a strictly supporting line at $u$ with slope $\beta$; i.e.,
\[
s(v) < s(u) + \beta(v-u) \ \mbox{ for all } v \not = u.
\]

\hspace{.25in} {\em (ii)} \mbox{{\em {\bf Partial equivalence.}}} $\:$ There exists
$\beta \in \R$ such that ${\cal E}^u \subset {\cal E}_\beta$ 
but ${\cal E}^u \not = {\cal E}_\beta$.  This is the case if and only
if $s$ has a nonstrictly supporting line at $u$ with slope $\beta$; i.e., 
\[
s(v) \leq s(u) + \beta(v-u) \ \mbox{ for all } v \in \r
\mbox{ with equality for some } v \not = u.
\]

\hspace{.25in} {\em (iii)} \mbox{{\em {\bf Nonequivalence.}}} $\:$ For all $\beta \in \R$,
$\eu \cap \ebeta = \emptyset$.  This is the case if and only if
$s$ has no supporting line at $u$; i.e., for any $\beta \in \R$
there exists $v$ such that $s(v) > s(u) + \beta(v-u)$.

{\em (b)} {\bf Canonical is always realized microcanonically.} $\:$
For $\nu \in \cp$ we define $\thi(\nu) = -\frac{1}{2}\lan \nu,\nu \ran$.
Then for any $\beta \in \R$ 
\[
\ebeta = \bigcup_{u \in \thi(\ebeta)} \eu.
\]
\end{thm}

We next relate ensemble equivalence and nonequivalence with concavity and support
properties of $s$ in the Curie-Weiss-Potts model.  
For $q = 3$ an explicit formula for $s$ is given in part (b) of Theorem \ref{thm:su}.  
If Conjecture \ref{conj:eu} is true, then the formula for $s$ given in part (c)
of Theorem \ref{thm:su} is also valid for $q \geq 4$.  
All the concavity and support features of $s$ 
are exhibited in Figure 1.  However, this figure is not the actual
graph of $s$ but a schematic graph that accentuates the shape of $s$
together with the intervals of strict concavity and nonconcavity of $s$.  

Concavity properties of $s$ are defined in reference to the double-Legendre-Fenchel
transform $s^{**}$, which can be characterized as the smallest concave, upper
semicontinuous function that satisfies $s^{**}(u) \geq s(u)$ for all $u \in \R$
\cite[Prop. A.2]{CosEllTouTur}.  For $u \in \mboxdoms$ we say that $s$ is concave
at $u$ if $s(u) = s^{**}(u)$ and that $s$ is not concave
at $u$ if $s(u) < s^{**}(u)$.  Also, we say that $s$ is strictly concave at 
$u \in \mboxdoms$ if 
$s$ has a strictly supporting line at $u$ and that $s$ is strictly concave
on a convex subset $A$ of $\mboxdoms$ if $s$ is strictly concave at each $u \in A$.

\begin{figure}[t]
\label{figure:entropy}
\begin{center}
\epsfig{file=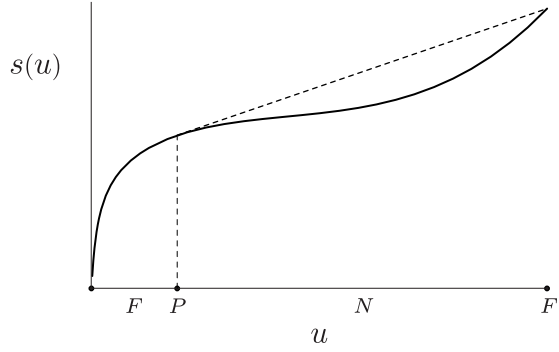}
\caption{ \small  Schematic graph of $s(u)$, showing the set
$F = (-\frac{1}{2}, u_0) \cup \{-\frac{1}{2q}\}$ of full ensemble
equivalence, the singleton set $P = \{u_0\}$ of partial equivalence, and the set
$N = (u_0,-\frac{1}{2q})$ of nonequivalence.
For $u \in F \cup P = (-\frac{1}{2}, u_0] \cup \{-\frac{1}{2}\}$,
$s(u) = s^{**}(u)$; for $u \in N$, $s(u) < s^{**}(u)$ and the graph of $s^{**}$
consists of the dotted line segment with slope $\beta_c$.  The slope of $s$
at $-\frac{1}{2}$ is $\infty$.}
\end{center}
\end{figure} 

According to Figure 1 and Theorem \ref{thm:equiv}, 
there exists $u_0 \in (-\frac{1}{2},-\frac{1}{2q})$ with the following
properties.
\begin{itemize}
\item $s$ is strictly concave on the interval
$(-\frac{1}{2},u_0)$ and at the point $-\frac{1}{2q}$.  Hence
for $u \in F = (-\frac{1}{2},u_0) \cup \{-\frac{1}{2q}\}$ 
the ensembles are fully equivalent [Thm.\ \ref{thm:equiv}(a)(i)]. 
In fact, for $u \in (-\frac{1}{2},u_0)$, $\eu = \ebeta$ with 
$\beta$ given by the thermodynamic formula $\beta = s'(u)$.
\item $s$ is concave
but not strictly concave at $u_0$ and
has a nonstrictly supporting line at $u_0$ that also touches the graph
of $s$ over the right hand endpoint $-\frac{1}{2q}$.  Hence for $u = u_0$ the
ensembles are partially equivalent in the sense that there exists
$\beta \in \r$ such that $\eu \subset \ebeta$ but $\eu \not = \ebeta$
[Thm.\ \ref{thm:equiv}(a)(ii)].  In fact, $\beta$ equals
the critical inverse temperature $\beta_c$ defined in (\ref{eqn:betac}).

\item $s$ is not concave on the interval 
$N = (u_0,-\frac{1}{2q})$ and has no supporting line 
at any $u \in N$ \cite[Thm.\ A.4(c)]{CosEllTouTur}.  Hence
for $u \in N$ the ensembles are nonequivalent in the sense that 
for all $\beta \in \r$, $\eu \cap \ebeta = \emptyset$ [Thm.\ \ref{thm:equiv}(a)(iii)].
\end{itemize}

We point out two additional features of Figure 1.
First, although $\eu \not = \emptyset$ for $u$ equal to the right hand endpoint $-\frac{1}{2}$ 
of $\mbox{dom} \, s$, we do not include this point in the set $F$ of full ensemble
equivalence.  Indeed, $s$ is not strictly concave at $-\frac{1}{2}$ because
there is no strictly supporting line at $-\frac{1}{2}$; as one can see in
(\ref{eqn:sprime}), the slope of $s$ at $-\frac{1}{2}$ is $\infty$.  Nevertheless, 
by introducing the limiting set
\[
{\cal E}_\infty = \{(1,0,\ldots,0), (0,1,\ldots,0), \ldots,(0,0,\ldots,1)\}
= \lim_{\beta \goto \infty} \ebeta,
\]
we can extend full ensemble equivalence to $u = -\frac{1}{2}$ since ${\cal E}^{-\frac{1}{2}}
= {\cal E}_\infty$.

Second, for $u$ in the interval $N$ of ensemble nonequivalence, 
the graph of $s^{**}$ is affine; this is depicted by the 
dotted line segment in Figure 1.  The slope of the affine portion
of the graph of $s^{**}$ equals the critical inverse temperature
$\beta_c$ defined in (\ref{eqn:betac}).  This can be proved using
concave-duality relationships involving $s^{**}$ and the canonical free energy.
The quantity $\beta_c$ also satisfies an equal-area property, first
observed by Maxwell \cite[p.\ 45]{Huang} and explained in the
context of another spin model in \cite[p.\ 535]{EllTouTur}.

The relationships stated in the three bulleted items above give valuable insight
into equivalence and nonequivalence of ensembles in the Curie-Weiss-Potts model.
These relationships are illustrated in Figure 2. 
In this figure we exhibit the graph of $s'$ and the sets
$\ebeta$ and $\eu$ in order to
compare the phase transitions in the two ensembles
and to understand the implications for ensemble equivalence and nonequivalence.
In order to accentuate properties of $s'$, $\ebeta$, and $\eu$ that
are related to ensemble equivalence and nonequivalence, we focus on $q = 8$.
In presenting the graph of $s'$ and the form of $\eu$, we assume that 
for $q = 8$ Conjecture \ref{conj:eu} is valid.  We then
appeal to part (c) of Theorem \ref{thm:su}, which gives
an explicit formula for $s$, and to part (d) of Theorem \ref{thm:eu}, which gives an explicit
formula for the elements of $\eu$.   The derivative $s'$, graphed in the 
top left plot in Figure 2, is given by
\be
\label{eqn:sprime}
s'(u) = \sqrt{\frac{q-1}{-2qu-1}} \left[ \log\!\left(1 + \sqrt{(q-1)(-2qu-1)} \,\, \right) 
- \log \!\left(1 - \sqrt{\frac{-2qu-1}{q-1}} \, \right) \right].
\ee

\begin{figure}[t]
\label{figure:phasediag}
\begin{center}
\epsfig{file=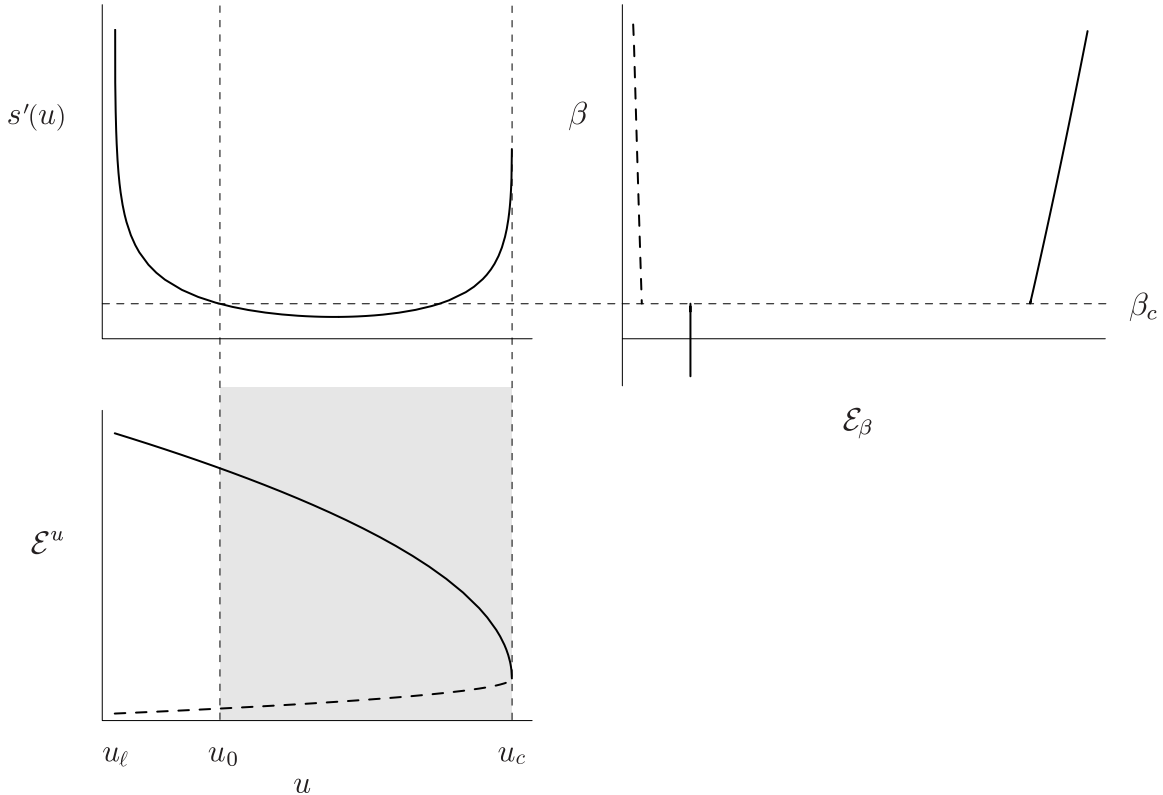}
\caption{\small  For $q = 8$ the top right plot shows $\ebeta$, the top left plot
the graph of $s'(u)$ for $u \in \mbox{dom} \, s =
[u_\ell,u_c] = [-\frac{1}{2},-\frac{1}{2q}]$,
and the bottom left plot $\eu$.  The discontinuous phase transition
at $\beta_c$ in the top right plot and the continuous phase
transition at $u_c$ in the bottom left plot imply
that the ensembles are nonequivalent for all $u \in N = (u_0,u_c)$.
On this interval $s$ is not concave and $s^{**}$ is affine with slope $\beta_c$.
The shaded area in the bottom left plot corresponds to the region
of nonequivalence of ensembles delineated by $u \in N$.}
\end{center}
\end{figure} 

The canonical phase diagram, given in the top right plot in Figure 2,
summarizes the description of $\ebeta$ given in Theorem \ref{thm:eb}
and shows the discontinuous phase transition exhibited
by this ensemble at $\beta_c = \frac{2(q-1)}{q-2} \log(q-1) = \frac{7}{3}\log 7$.
The solid line in this plot for $\beta < \beta_c$ represents the common
value $\frac{1}{8}$ of each of the components of $\rho$, which is the unique 
phase for $\beta < \beta_c$. 
For $\beta > \beta_c$ there are eight phases given by $\nu^1(\beta)$ together
with the vectors $\nu^i(\beta)$ obtained by interchanging the first and $i$th components of $\nu^1(\beta)$. 
Finally, for $\beta = \beta_c$ there are nine phases consisting of $\rho$ and
the vectors $\nu^i(\beta_c)$ for $i = 1, 2, \ldots, 8$.
The solid and dashed curves in the top right plot in Figure 2
show the first component and the last seven, equal components of
$\nu^1(\beta)$ for $\beta \in [\beta_c,\infty)$.  The first component
is a strictly increasing function
equal to $\frac{7}{8}$ for $\beta = \beta_c$ and increasing to $1$ as
$\beta \goto \infty$ while the last seven, equal components 
are strictly decreasing functions equal to $\frac{1}{56}$ for
$\beta = \beta_c$ and decreasing to $0$ as $\beta \goto \infty$.  

The microcanonical phase diagram, given in the bottom left plot in Figure 2,
summarizes the description of $\eu$ given in Theorem \ref{thm:eu}
and shows the continuous phase transition exhibited
by this ensemble as $u$ decreases from the maximum value
$u_c = -\frac{1}{2q} = -\frac{1}{16}$.
The single phase $\rho$ for $u = -\frac{1}{16}$ is represented by the point lying
over this value of $u$.  For $u \in [-\frac{1}{2}, -\frac{1}{16})$ 
there are eight phases given by $\nu^1(u)$ together
with the vectors $\nu^i(u)$ obtained by interchanging the 
first and $i$th components of $\nu^1(u)$.  
The solid and dashed curves in the bottom left plot in Figure 2 show the
first component $a(u)$ and the last seven, equal components
$b(u)$ of $\nu^1(u)$ for $u \in [-\frac{1}{2}, -\frac{1}{16})$.
The first component is a strictly increasing function of $-u$ equal
to $\frac{1}{8}$ for $u = -\frac{1}{16}$ and increasing to 1 as $u \goto -\frac{1}{2}$,
while the last seven, equal components are strictly decreasing functions 
of $-u$ equal
to $\frac{1}{8}$ for $u = -\frac{1}{16}$ and decreasing to 0 as
$u \goto -\frac{1}{2}$.

The different nature of the two phase transitions --- discontinuous in the 
canonical ensemble versus continuous in the microcanonical ensemble --- implies
that the two ensembles are not fully equivalent 
for all values of $u$.  By necessity, the set $\ebeta$ of canonical equilibrium macrostates
must omit a set of microcanonical equilibrium macrostates.  Further
details concerning ensemble equivalence and nonequivalence can be
seen by examining the graph of $s'$, given in the top left plot of Figure 2.
This graph, which is the bridge between the canonical and microcanonical
phase diagrams, shows that
$s'$ is strictly decreasing on the interval $\mbox{int} \, F = 
(-\frac{1}{2},u_0)$, which is the interior of the set $F$ 
of full ensemble equivalence.  
The critical value $\beta_c$ equals the slope of the affine portion of 
the graph of $s^{**}$
over the interval $N = (u_0,-\frac{1}{2q})$ of ensemble nonequivalence. 
This affine portion is represented in the top left plot of Figure 2
by the horizontal dashed line at $\beta_c$. 
 
Figure 2 exhibits the full equivalence of ensembles that holds for $u \in \mbox{int} \, F = 
(-\frac{1}{2},u_0)$ [Thm.\ \ref{thm:euebeta}(a)].  For $u$ in this interval
the solid and dashed curves representing the components of $\nu^1(u) \in \eu$
can be put in one-to-one correspondence with the solid and dashed
curves representing the same two components of $\nu^1(\beta) \in \ebeta$
for $\beta \in (\beta_c,\infty)$.  The values of $u$ and $\beta$ are 
related by $s'(u) = \beta$.
Full equivalence of ensembles also holds for $u = -\frac{1}{2q} \in F$,
the right-hand endpoint of the interval on which $s$ is finite.  The solid vertical line
in the top right plot for $\beta < \beta_c$, which represents the unique phase
$\rho$, is collapsed to the single point representing the unique phase
$\rho$ for $u = -\frac{1}{2q}$ in the bottom left plot.  
This collapse shows that the canonical notion of temperature is somewhat
ill-defined at $u = -\frac{1}{2q}$ since lowering $\beta$ down to $\beta_c$ 
changes neither the equilibrium macrostate $\rho$ nor 
the associated mean energy $u$.  
This feature of the Curie-Weiss-Potts model is not present, for example,
in the mean-field Blume-Emery-Griffiths spin model, which also exhibits nonequivalence
of ensembles \cite{EllTouTur}.  

By comparing the top right and bottom left plots, we see that
the elements of $\eu$ cease to be related to those of $\ebeta$ 
for $u \in N = (u_0,-\frac{1}{2q})$, which is the interval on 
which $s$ is not concave.  For any mean-energy value $u$ in this interval 
no $\nu \in \ebeta$ exists that can be put in
correspondence with an equivalent equilibrium empirical vector contained in
$\mathcal{E}^u$. 
Thus, although the equilibrium macrostates 
corresponding to $u\in N$ are characterized by a well defined value
of the mean energy, it is impossible to assign an inverse temperature $\beta$ to those
macrostates from the viewpoint of the canonical ensemble. In other words,
the canonical ensemble is blind to all mean-energy values $u$ contained in
the interval $N$ of nonconcavity of $s$.  This is closely related to the
presence of the discontinuous phase transition seen in the canonical
ensemble.

The quantity $u_0$ defined in (\ref{eqn:u0}) plays a central
role in the analysis of phase transitions and ensemble equivalence
in the Curie-Weiss-Potts model.  First, as we saw in our discussion 
of Figure 1, $u_0$ separates
the interval $(-\frac{1}{2},u_0)$ of full ensemble equivalence
from the interval $(u_0,-\frac{1}{2q})$ of nonequivalence.
Second, part (a) of Lemma \ref{lemma:ubeta} shows that $u_0$
equals the limiting mean energy $\tilde{H}(\nu^1(\beta_c))$
in the canonical equilibrium macrostate $\nu^1(\beta)$ as $\beta \goto (\beta_c)^+$.
In Figures 3 and 4 we present for $q = 3$ a third, geometric interpretation of 
$u_0$ that is also related to nonequivalence of ensembles.

Before explaining this third, geometric interpretation of $u_0$, we
recall that according to part (a) of Theorem \ref{thm:eu},
$\eu$ is nonempty, or equivalently the constraint set in (\ref{eqn:cu})
is nonempty, if and only if $u \in [-\frac{1}{2},-\frac{1}{2q}] = 
[-\frac{1}{2},-\frac{1}{6}]$.  Geometrically,
the energy constraint $\tilde{H}(\nu) = -\frac{1}{2}\lan \nu,\nu \ran = u$
corresponds to the sphere in $\R^3$ with center 0 and radius $\sqrt{-2u}$.
This sphere
intersects the set $\cp$ of probability vectors
if and only if $u \in [-\frac{1}{2},-\frac{1}{6}]$.
For $u = -\frac{1}{6}$, the sphere is tangent to $\cp$ at the unique point $\rho$
while for $u = -\frac{1}{2}$, the hypersphere intersects $\cp$ at the $q$ unit-coordinate
vectors.  The intersection
of the sphere and $\cp$ undergoes a phase transition at $u_0$
in the following sense.  For $u \in [u_0,-\frac{1}{6})$
the sphere intersects $\cp$ in a circle
 while for $u \in [-\frac{1}{2},u_0)$,
the sphere intersects $\cp$ in a proper subset of a circle; the complement of this subset
lies outside the nonnegative octant of $\R^3$.  For $u = u_0 = -\frac{1}{4}$, 
the circle of intersection
is maximal and is tangent to the boundary of $\cp$. 

The set $\ebeta$ of canonical equilibrium macrostates for $q = 3$ is represented
in Figure 3.  
In this figure the maximal circle of intersection corresponding to $u = u_0 = -\frac{1}{4}$ 
is shown together with the vector $\rho$ at its center; the points $A$, 
$B$, and $C$ representing
the respective unit-coordinate vectors $(1,0,0)$, $(0,1,0)$, and $(0,0,1)$; and the
points $A_c$, $B_c$, and $C_c$ representing the respective equilibrium macrostates
$\nu^1(\beta_c)$, $\nu^2(\beta_c)$, and $\nu^3(\beta_c)$.  These three macrostates
lie on the maximal circle of intersection since $\tilde{H}(\nu^1(\beta_c)) = u_0$
[Lem.\ \ref{lemma:ubeta}(b)].  For $\beta > \beta_c$ all $\nu \in \ebeta$
have two equal components, and as $\beta \goto \infty$ these vectors
converge to the unit-coordinate vectors $A$, $B$, and $C$.  Hence for $\beta > \beta_c$
the equilibrium macrostates $\nu^1(\beta)$, $\nu^2(\beta)$, and $\nu^3(\beta)$
are represented by the open line segments $\overline{A_cA}$,
$\overline{B_cB}$, and $\overline{C_cC}$.

\begin{figure}[t]
\label{figure:ebeta}
\begin{center}
\epsfig{file=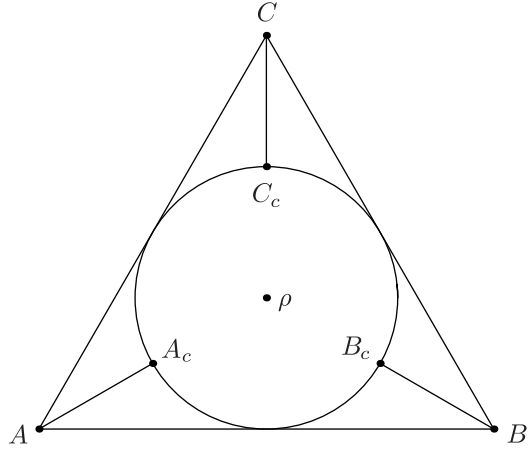}
\caption{ \small Graphical representation of the set 
$\ebeta$ of canonical equilibrium macrostates for $q = 3$ showing the maximal circle of intersection
corresponding to $u = u_0$; the vector $\rho$\,; the unit-coordinate vectors $A$, 
$B$, and $C$; and
the macrostates $A_c = \nu^1(\beta_c)$, $B_c = \nu^2(\beta_c)$, 
and $C_c = \nu^3(\beta_c)$.  The line segments $\overline{A_cA}$,
$\overline{B_cB}$, and $\overline{C_cC}$ represent 
the elements of $\ebeta$ for $\beta > \beta_c$.}
\end{center}
\end{figure}

\begin{figure}[t]
\label{figure:eu}
\begin{center}
\epsfig{file=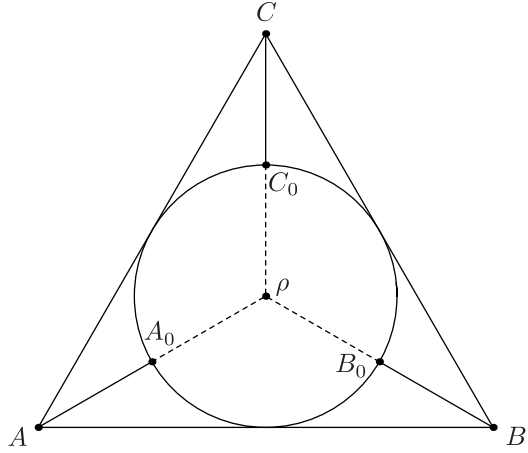}
\caption{ \small Graphical representation of the set 
$\eu$ of microcanonical equilibrium macrostates for $q = 3$ showing the maximal circle of intersection
corresponding to $u = u_0$; the vector $\rho$\,; the unit-coordinate vectors $A$,
$B$, and $C$; and the macrostates $A_0 = \nu^1(u_0)$, $B_0 = \nu^2(u_0)$, 
and $C_0 = \nu^3(u_0)$.  The solid-line segments $\overline{A_0A}$,
$\overline{B_0B}$, and $\overline{C_0C}$ represent 
the elements of $\eu$ that are realized canonically.  The dashed-line segments
$\overline{\rho A_0}$, $\overline{\rho B_0}$,
and $\overline{\rho C_0}$ represent the elements of $\eu$ that are not realized canonically.}
\end{center}
\end{figure} 

The set $\eu$ of microcanonical equilibrium macrostates for $q = 3$ is represented
in Figure 4.  In this figure the maximal circle of intersection 
corresponding to $u = u_0 = -\frac{1}{4}$ is shown together
with the vector $\rho$ at its center; the points $A$, $B$, and $C$ representing
the unit-coordinate vectors; and the points $A_0$, $B_0$, and $C_0$ 
representing the respective equilibrium macrostates
$\nu^1(u_0)$, $\nu^2(u_0)$, and $\nu^3(u_0)$.
For $u \in (-\frac{1}{2},-\frac{1}{6})$ all $\nu \in \eu$
have two equal components, and as $u \goto -\frac{1}{2}$ they converge
to the unit coordinate vectors $A$, $B$, and $C$.  Hence for 
$u \in (-\frac{1}{2},-\frac{1}{6})$
the equilibrium macrostates $\nu^1(u)$, $\nu^2(u)$, and $\nu^3(u)$
are represented by the open line segments $\overline{\rho A}$,
$\overline{\rho B}$, and $\overline{\rho C}$.
As we saw in the preceding section,
for each $u \in (-\frac{1}{2}, -\frac{1}{6})$ the macrostates 
$\nu^1(u)$, $\nu^2(u)$, and $\nu^3(u)$ lie on the intersection of the sphere
of radius $\sqrt{-2u}$ with $\cp$.  In particular, $A_0 = \nu^1(u_0)$, 
$B_0 = \nu^2(u_0)$, and $C_0 = \nu^3(u_0)$
lie on the maximal circle of intersection.  

The distinguishing feature of Figure 4 is
the three open dashed-line segments $\overline{\rho A_0}$, $\overline{\rho B_0}$, 
and $\overline{\rho C_0}$
representing the elements of $\eu$ that are not realized canonically;
namely, $\nu^1(u)$, $\nu^2(u)$, and $\nu^3(u)$ for $u \in (u_0,-\frac{1}{6})$. 
The three half open solid-line segments $\overline{A_0A}$, $\overline{B_0B}$,
and $\overline{C_0C}$
represent the elements of $\eu$ that are realized canonically;
namely, $\nu^1(u)$, $\nu^2(u)$, and $\nu^3(u)$ for $u \in (-\frac{1}{2},u_0]$. 
For each such $u$ the value of $\beta$ for which $\eu = \ebeta$
is determined by the equation $\tilde{H}(\nu^1(\beta)) = u$ [Thm.\ \ref{thm:euebeta}(a)].  
Thus in Figure 3 the corresponding
elements of $\ebeta$ 
lie on the intersection of the sphere of radius $\sqrt{-2u}$ and $\cp$.

This completes
our discussion of equivalence and nonequivalence of ensembles.
In the next section we will prove a number of statements concerning
ensemble equivalence and nonequivalence that have been determined graphically.

\section{Proofs of Equivalence and Nonequivalence of Ensembles}
\beginsec

Using the general results of \cite{EHT}, we stated in the preceding section
the equivalence and nonequivalence relationships that exist between $\eu$
and $\ebeta$ and verified these relationships using the plots of
these sets for $q=8$ given in Figure 2. Our purpose in the present section is
to prove these relationships using mapping properties of the mean energy
function $u(\beta)$ defined for $\beta 
\not = \beta_c$ by 
\be
\label{eqn:ubeta2}
u(\beta) = \left\{ \begin{array}{ll} \thi(\rho) = -\frac{1}{2q}
 & \ \mbox{ for }  \: \beta < \beta_c,  \\
\thi(\nu^1(\beta)) = -\frac{1}{2}\lan \nu^1(\beta),
\nu^1(\beta) \ran  & \ \mbox{ for }  \: \beta > \beta_c.
\end{array}
\right.  
\ee
Here $\nu^1(\beta)$ is the unique canonical equilibrium macrostate modulo permutations
for $\beta > \beta_c$ [Thm.\ \ref{thm:eb}].
According to the next lemma, for $\beta > \beta_c$, $u(\beta)$ is continuous
and strictly decreasing and $u(\beta) < -\frac{1}{2q}$, which equals 
the mean energy for $\beta < \beta_c$.  It follows that as $\beta$ increases through
$\beta_c$, $u(\beta)$ is discontinuous, jumping down from $-\frac{1}{2q}$ to
$\thi(\nu^1(\beta))$.  This discontinuity in $u(\beta)$ mirrors in a natural way
the discontinuity in $\ebeta$ as $\beta$ increases through $\beta_c$.

\begin{lemma}\per
\label{lemma:ubeta}
For $\beta \in [\beta_c,\infty)$ we define $\nu^1(\beta)$ as 
in part {\em (b)} of Theorem {\em \ref{thm:eb}} and we define
\be
\label{eqn:u0}
u_0 = \frac{-q^2 + 3q - 3}{2q(q-1)},
\ee 
The following conclusions hold.

{\em (a)} $-\frac{1}{2} < u_0 < -\frac{1}{2q}$ and
$\lim_{\beta \goto (\beta_c)^+} u(\beta) = \thi(\nu^1(\beta_c)) = u_0$. 

{\em (b)} The function mapping 
\[
\beta \in (\beta_c,\infty) \mapsto u(\beta) = \thi(\nu^1(\beta)) = - \ts \frac{1}{2}
\lan \nu^1(\beta),\nu^1(\beta) \ran
\]
is a strictly decreasing, differentiable bijection onto the interval $(-\frac{1}{2},u_0)$.  
\end{lemma}

\noi
{\bf Proof.}  (a) The inequalities involving $u_0$ 
follow immediately from the inequality $q \geq 3$.  
The relationship $\thi(\nu^1(\beta_c)) = u_0$ is easily
determined using the explicit form of $\nu^1(\beta_c)$
given in (\ref{eqn:nu1betac}).  That 
$\lim_{\beta \goto (\beta_c)^+} u(\beta) = \thi(\nu^1(\beta_c))$
follows from the definition of $u(\beta)$ and the continuity of $\nu^1(\beta)$ for
$\beta > \beta_c$.

(b) For $\beta \in (\beta_c,\infty)$ we use the formula for $\nu^1(\beta)$ given in part (b)
of Theorem \ref{thm:eb} to calculate
\[
u(\beta) = -\frac{1}{2}\left(\frac{[1+(q-1)w(\beta)]^2}{q^2} +
(q-1)\frac{[1-w(\beta)]^2}{q^2}\right).
\]
Since $w(\beta)$ is positive, strictly increasing, and differentiable
for $\beta \in (\beta_c,\infty)$ [Thm.\ \ref{thm:eb}(a)] and since 
\[
u'(\beta)= \frac{-(q-1)w(\beta)w'(\beta)}{q} < 0 \ \mbox{ for } \beta
\in (\beta_{c}, \infty),
\]
$u(\beta)$ is strictly decreasing for $\beta \in (\beta_{c},\infty)$.
In addition, since $\lim_{\beta \goto \infty} w(\beta) = 1$ [Thm.\ \ref{thm:eb}(a)], 
we have $\lim_{\beta \goto \infty} u(\beta) = -\frac{1}{2}$,
and by part (a) of this lemma
\[
\lim_{\beta \goto (\beta_c)^+} u(\beta) = \lim_{\beta \goto (\beta_c)^+} 
\thi(\nu(\beta_c)) = u_0.
\]
It follows that the function mapping $\beta \in (\beta_c,\infty) \mapsto u(\beta)$
is a strictly decreasing, differentiable bijection onto the interval $(-\frac{1}{2},-\frac{1}{2}u_0)$. 
This completes the proof of part (b).  \ink

\skp
Mapping properties of $u(\beta)$ play an important role in the next theorem, in 
which we prove that the sets $F$, $P$, and $N$ defined in (\ref{eqn:fpn})
correspond to 
full equivalence, partial equivalence, and nonequivalence of ensembles.
For $u \in F$
we consider three subcases in order to indicate the value of
$\beta$ for which $\eu = \ebeta$; for $u \in \mbox{int} \, F
= (-\frac{1}{2},u_0)$, $\beta$ and $u$ are related
by $\beta = s'(u)$ and $u = u(\beta)$.    Part (c) shows an interesting
degeneracy in the equivalence-of-ensemble picture, the set $\eu$
for $u = -\frac{1}{2q}$ corresponding to all $\ebeta$ for $\beta 
< \beta_c$.  This is related to the fact that for all such values
of $\beta$, $\ebeta = \{\rho\}$ and thus the mean energy $u(\beta)$
equals $-\frac{1}{2q}$.

\begin{theorem}\per
\label{thm:euebeta}   
We define $s(u)$ in {\em (\ref{eqn:microentropy})},
$u(\beta)$ in {\em (\ref{eqn:ubeta2})}, $\ebeta$ in {\em (\ref{eqn:canonmacro})},
and $\eu$ in {\em (\ref{eqn:micromacro})}.  We also define $\beta_c$
in {\em (\ref{eqn:betac})} and $u_0$ in {\em (\ref{eqn:u0})}.
The sets
\be
\label{eqn:fpn}
F = (\ts-\frac{1}{2},u_0) \cup \{\ts-\frac{1}{2q}\}, \ 
P = \{\ts u_0\}, \ \mbox{ and } N = (\ts-\frac{1}{2}u_0, -\frac{1}{2q})
\ee
have the following properties.

\vspace{.1in}

{\em (a)} {\em {\bf Full equivalence on int}} \boldmath $F$. \unboldmath 
For $u \in \mbox{{\em int}} \, F = 
(-\frac{1}{2},u_0)$, there exists a unique $\beta 
\in (\beta_c,\infty)$ such that ${\cal{E}}^u={\cal{E}}_\beta$;
$\beta$ satisfies $u(\beta) = \thi(\nu^1(\beta)) = u$.

\vspace{.1in}  

{\em (b)}  For $u \in \mbox{{\em int}} \, F = (-\frac{1}{2},u_0)$, $s$
is differentiable.  The values $u$ and $\beta$ for which $\eu = \ebeta$
in part {\em (a)} are also related by the thermodynamic formula $s'(u) = \beta$.

\vspace{.1in}

{\em (c)} {\em {\bf Full equivalence at}} \boldmath $-\frac{1}{2q}$. \unboldmath 
For $u = -\frac{1}{2q} \in F$, ${\cal E}^{-\frac{1}{2q}} = {\cal E}_\beta$ for any
$\beta < \beta_c$.

\vspace{.1in}

{\em (d)} {\em {\bf Partial equivalence on}} \boldmath $P$. \unboldmath
For $u \in P = \{u_0\}$, ${\cal{E}}^{u_0} \subset {\cal{E}}_{\beta_c}$
but ${\cal{E}}^{u_0} \not= {\cal{E}}_{\beta_c}$.  In fact,
${\cal{E}}_{\beta_c} = {\cal{E}}^{u_0} \cup {\cal E}^{-\frac{1}{2q}}$.

\vspace{.1in}

{\em (e)} {\em {\bf Nonequivalence on}} \boldmath $N$. \unboldmath
For any $u \in N = (u_0, -\frac{1}{2q})$, ${\cal{E}}^u \cap
{\cal{E}}_\beta = \emptyset$ for all $\beta \in \R$.  
\end{theorem}

In reference to the properties of $s$ given in part (b), 
one can show that the function mapping $u \in (-\frac{1}{2},u_0) \mapsto
s'(u)$ is a strictly decreasing, differentiable bijection onto 
the interval $(\beta_c,\infty)$ and that this bijection is the
inverse of the bijection mapping $\beta \in (\beta_c,\infty) \mapsto u(\beta)$. 

Before we prove the theorem, it is instructive to compare its assertions
with those in Theorem \ref{thm:equiv}, which formulates
ensemble equivalence and nonequivalence in terms of support properties of $s$.
These support properties can be seen in the schematic plot of the 
the graph of $s$ in Figure 1.  We start with part (a)
of Theorem \ref{thm:euebeta}, which states that for any $u \in \mbox{int} \, F
= (-\frac{1}{2},u_0)$ there exists a unique 
$\beta \in (\beta_c,\infty)$ such that $\eu = \ebeta$.  As promised in part 
(a)(i) of Theorem \ref{thm:equiv}, this $\beta$ is the slope of a strictly
supporting line to the graph of $s$ at $u$.  
The situation that holds when $u = -\frac{1}{2q}$ [Thm. 
\ref{thm:euebeta}(c)] is also consistent
with part (a)(i) of Theorem \ref{thm:equiv}.  For this value of $u$, 
which is the isolated point of the set $F$ of full equivalence, there
exist infinitely many strictly supporting lines to the graph of $s$,
the possible slopes of which are all $\beta \in (-\infty,\beta_c)$.
On the other hand, when $u = u_0$, which is the only value lying
in the set $P$ of partial equivalence, we have 
${\cal{E}}^{u_0} \subset {\cal{E}}_{\beta_c}$
but ${\cal{E}}^{u_0} \not= {\cal{E}}_{\beta_c}$ [Thm.\ \ref{thm:euebeta}(d)].
In combination with part (a)(ii) of Theorem \ref{thm:equiv}, 
it follows that there exists a nonstrictly 
suppporting line at $u$ with slope $\beta_c$.  
Finally, for $u \in N = (u_0, -\frac{1}{2q})$, we have
$\eu \cap \ebeta = \emptyset$ for all $\beta \in \R$ [Thm.\ 
\ref{thm:euebeta}(e)].  In accordance
with part (a)(iii) of Theorem \ref{thm:equiv}, $s$ has no supporting line at 
any $u \in N$, and by Theorem A.4 in \cite{CosEllTouTur} $s$ is not
concave at any $u \in N$.

\skp
\noi
{\bf Proof of Theorem \ref{thm:euebeta}.}  
(a) For $\beta > \beta_c$ part (b) of 
Theorem \ref{thm:eb} and part (b) of Theorem \ref{thm:equiv} imply that 
\[
{\cal{E}}_{\beta} = \{\nu^{1}(\beta),\ldots,\nu^{q}(\beta)\} 
= \bigcup_{u \in \tilde{H}({\cal{E}}_\beta)} {\cal{E}}^{u}.
\]
The symmetry of $\thi$ with respect
to permutations implies that $\tilde{H}({\cal{E}}_\beta) = \{
\tilde{H}(\nu^{1}(\beta)) \}$.  Thus for any $\beta > \beta_c$
\be
\label{eqn:thishelps}
{\cal{E}}_\beta={\cal{E}}^{\tilde{H}(\nu^{1}(\beta)) }.
\ee
Since for any $u \in \mbox{int} \, F = (-\frac{1}{2}, u_0)$ there exists
a unique $\beta \in (\beta_c,\infty)$ satisfying $u(\beta) = \thi(\nu^1(\beta)) = u$ 
[Lem.\ \ref{lemma:ubeta}(b)], it follows that $\eu = \ebeta$.  

(b) According to part (b) of Theorem \ref{thm:rigorq4}, $s$ 
is differentiable at all $u \in \mbox{int} \, F$.  
Since $s = s^{**}$ in a neighborhood of each such $u$, 
part (a) of Theorem A.3 in \cite{CosEllTouTur} implies that $s'(u) = \beta$.

(c) By (\ref{eqn:onepoint}) and part (b) of Theorem \ref{thm:eb}
\be
\label{eqn:e12q}
{\cal E}^{-\frac{1}{2q}} = \{\rho\} = \ebeta \ \mbox{ for any } \beta < \beta_c.
\ee

(d) By part (b) of Theorem \ref{thm:eb}, symmetry, and part (a) of Lemma
\ref{lemma:ubeta}
\[
\tilde{H}({\cal{E}}_{\beta_c}) = \{\thi(\rho), \thi(\nu^1(\beta_c))\}
= \{\ts-\frac{1}{2q},u_0\}.
\]
Hence by (\ref{eqn:thishelps}) and (\ref{eqn:e12q})
\[
{\cal E}_{\beta_c} = \bigcup_{u \in \tilde{H}({\cal{E}}_{\beta_c})}
{\cal E}^{u}= {\cal E}^{-\frac{1}{2q}} \cup {\cal E}^{u_0} = 
\{\rho\} \cup {\cal E}^{u_0}. 
\]
However, $\rho \notin {\cal{E}}^{u_0}$ since $\rho$ does not satisfy the constraint
$\tilde{H}(\rho) = u_0$.  It follows that ${\cal{E}}^{u_0} \subset {\cal{E}}_{\beta_c}$
but that ${\cal{E}}^{u_0} \not = {\cal{E}}_{\beta_c}$.  

(e) If $u \in N$, then $u \notin 
(-\frac{1}{2},u_0)$, and so by
part (a) of Lemma \ref{lemma:ubeta} $u \not = \thi(\nu^1(\beta))$ 
for any $\beta \in (\beta_c,\infty)$.   Since by (\ref{eqn:thishelps})
${\cal{E}}_{\beta}={\cal{E}}^{\tilde{H}(\nu^{1}(\beta))}$ for all
$\beta > \beta_c$, it follows that for all $\beta > \beta_c$ 
\[
{\cal{E}}^u \cap {\cal{E}}^{\tilde{H}(\nu^{1}(\beta))} = \emptyset
\] 
and thus that ${\cal{E}}^u \cap {\cal{E}}_\beta = \emptyset$.
For any $\beta < \beta_c$ (\ref{eqn:e12q}) states
that ${\cal E}_\beta = {\cal E}^{-\frac{1}{2q}} = \{\rho\}$. 
Since $u \in N$, we have $u \not = -\frac{1}{2q}$ and thus
${\cal E}^{-\frac{1}{2q}} \cap {\cal E}^u =
\emptyset$.  It follows that ${\cal{E}}^u \cap {\cal{E}}_\beta = \emptyset$ for
any $\beta < \beta_c$.
Finally, for $\beta = \beta_c$ part (b) of Theorem \ref{thm:eb} states that 
${\cal E}_{\beta_c} = \{\rho, \nu^1(\beta_c), \ldots, \nu^q(\beta_c)\}$. 
However, since $\tilde{H}(\rho) = -\frac{1}{2q} \notin N$ and
$\tilde{H}(\nu^i(\beta_c)) = u_0 \notin N$, none of the vectors in ${\cal E}_{\beta_c}$
satisfies the constraint $\tilde{H}(\nu)=u$ .  Thus ${\cal{E}}^u \cap
{\cal{E}}_{\beta_c} = \emptyset$.  We have proved ${\cal{E}}^u \cap
{\cal{E}}_\beta = \emptyset$ for all $\beta \in \R$.  The proof
of the theorem is complete.  \ \ink

\skp
We end this section by showing that for arbitrary $q \geq 4$ and $u$ in the equivalence sets 
$F \cup P = (-\frac{1}{2},u_0] \cup \{-\frac{1}{2q}\}$ 
the formulas for $\eu$ and $s(u)$ given in part (d) of Theorem \ref{thm:eu}
and part (c) of Theorem \ref{thm:su} are rigorously true. 
Our strategy is to use the
equivalence of the microcanonical and canonical ensembles for 
$u \in F \cup P$ and
the fact that the form of $\ebeta$ is known exactly for all $\beta$. 
Thus, we translate the form of $\nu \in \ebeta$,
as given in part (b) of Theorem \ref{thm:eb},  
into the form of $\nu \in \eu$ for $u \in F \cup P$.  
For $\beta \in [\beta_c,\infty)$, the last $q-1$ components
of $\nu^1(\beta) \in \ebeta$  are given by
\be
\label{eqn:components}
\nu^1_j(\beta) = \frac{1 - w(\beta)}{q},
\ee
and these components are not equal to the first component.  
Since for each $u \in F \cup P$
there exists $\beta \in [\beta_c,\infty]$
such that either ${\cal{E}}^u = {\cal{E}}_\beta$ or $\eu \subset \ebeta$,
it follows that modulo permutations
all $\nu \in {\cal{E}}^u$ have their last $q-1$
components equal to each other.  That is, modulo permutations there exist numbers $a$ and $b$
in $[0,1]$ such that $\nu = (a, b, \ldots, b)$.  
The possible values of $a$ and $b$ are easily determined by considering the 
constraints satisfied by $\nu \in \eu$.  These constraints are
\[
\label{eqn:constraints}
 a + (q-1)b = 1 \ \mbox{ and } \ a^{2} + (q-1)b^{2} = -2u.
\]
The two solutions of these equations are
\[
a_1 = \frac{1-\sqrt{(q-1)(-2qu -1)} }{q}, \ \ 
b_1 = \frac{q-1+\sqrt{(q-1)(-2qu-1)}}{(q-1)q}
\]
and
\[
a_2 = \frac{1+\sqrt{(q-1)(-2qu-1)} }{q}, \ \ b_2 = \frac{q-1-\sqrt{(q-1)(-2qu-1)}
}{(q-1)q}.
\]
Of the two values $b_1$ and $b_2$, only $b_2$ has the form given
in (\ref{eqn:components}) with 
\[
w(\beta) = \frac{\sqrt{(q-1)(-2qu-1)}}{q-1} \in [0,1].
\]   
We conclude that modulo permutations each $\nu \in \eu$ has the form 
$(a_2,b_2,\ldots,b_2)$, in which the last $q-1$ components all equal $b_2$.
This coincides with the formula for $\nu^1(u)$ given in part (d)
of Theorem \ref{thm:eu}, which in turn gives the explicit formula for
$s(u)$ in part (c) of Theorem \ref{thm:su}.  This information is summarized
in part (a) of the next theorem.
The differentiability of $s$ on $\mbox{int} \, F$, which is stated in part (b),
is an immediate consequence of the explicit formula for $s(u)$.

\begin{thm}\per
\label{thm:rigorq4}
We define $u_0$ in {\em (\ref{eqn:u0})}.  The following conclusions hold.

{\em (a)}  For arbitrary $q \geq 4$ and $u$ in the equivalence sets
$F \cup P = (-\frac{1}{2},u_0] \cup \{-\frac{1}{2q}\}$ 
the formulas for $\eu$ and $s(u)$ given in part {\em (d)} 
of Theorem {\em \ref{thm:eu}}
and part {\em (c)} of Theorem {\em \ref{thm:su}} are rigorously true.

{\em (b)} For arbitrary $q \geq 4$,
$s$ is differentiable on the interval $\mbox{{\em int}} \, F = (-\frac{1}{2},u_0)$
and $s'(u)$ is given by {\em (\ref{eqn:sprime})}.
\end{thm}

\appendix
\renewcommand{\thesection}{\Alph{section}}
\renewcommand{\theequation}
{\Alph{section}.\arabic{equation}}
\renewcommand{\thedefn}
{\Alph{section}.\arabic{defn}}
\renewcommand{\theass}
{\Alph{section}.\arabic{ass}}

\section{Two Related Maximization Problems}
\beginsec

Theorem \ref{thm:relatedmin} is a new result on the maximum points of certain
functions related by convex duality.
It is formulated for a finite, differentiable,
convex function $F$ on $\rsigma$ and its Legendre-Fenchel transform
\[
F^*(z) = \sup_{x \in \rsigma}\{\lan x,z\ran - F(x)\}.
\]
With only minor changes in notation the theorem is also valid for a finite,
Gateaux-differentiable, convex function on a Hilbert space.

Theorem \ref{thm:relatedmin} will be applied in Appendix B
to prove that for $\beta > 0$, $\ebeta$ has the form
given in part (b) of Theorem \ref{thm:eb}.
Another application of Theorem \ref{thm:relatedmin} is given
in Proposition 3.4 in \cite{EllOttTou}.  It is used there to determine
the form of the set of canonical equilibrium macrostates for another
important spin system known as the mean-field Blume-Emery-Griffiths model.

\begin{theorem}\per
\label{thm:relatedmin}  Let $\sigma$ be a positive
integer and $F$ a finite, differentiable, convex function mapping $\R^\sigma$
into $\R$.
Assume that $\sup_{z \in \rsigma}\{F(z)-\frac{1}{2}\|z\|^2 \} < \infty$ and that
$F(z)-\frac{1}{2}\|z\|^2 $ attains its supremum.  The following conclusions hold.

{\em (a)} $\sup \limits_{z \in \R^\sigma} \{F(z)-\frac{1}{2}\|z\|^2\} = \sup \limits_{z \in
\mbox{{\scriptsize \rm dom}} \, F^*} \{\frac{1}{2}\|z\|^2 -F^*(z)\}$.

{\em (b)} \rule{0in}{.2in} $\frac{1}{2}\|z\|^2  - F^*(z)$ attains its supremum on $\mbox{{\em dom}} \, F^*$.

{\em (c)} \rule{0in}{.2in} The global maximum points of $F(z)-\frac{1}{2}\|z\|^2$ coincide with the
global maximum points of $\frac{1}{2}\|z\|^2 - F^*(z)$.
\end{theorem}

\noi
\textbf{Proof.} We define the subdifferential of $F^*$ at $z_0 \in \rsigma$
by
\[
\partial F^*(z_0) = \{y \in \rsigma : F^*(z) \geq F^*(z_0) + \lan y,z-z_0 \ran 
\mbox{ for all } z \in \rsigma\}.
\]
We also define the domain of $\partial F^*$ to be the set of $z_0 \in \rsigma$ 
for which $\partial F^*(z_0) \not = \emptyset$.
The proof of the theorem uses three properties of Legendre-Fenchel transforms.  
\begin{enumerate}
\item $F^*$ is a convex, lower semicontinuous function mapping $\rsigma$ into
$\r \cup \{\infty\}$, and
for all $z \in \rsigma$, $F^{**}(z) = (F^*)^*(z)$ equals 
$F(z)$ \cite[Thm.\ VI.5.3(a),(e)]{Ellis}.
\item If for some $z_0 \in \rsigma$ and $z \in \rsigma$ we have $z = \nabla F(z_0)$, then 
$F(z_0) + F^*(z) = \lan z_0, z \ran$, and so 
$z \in \mbox{dom} \, F^*$.
In particular, if $z = z_0$, then $z_0 \in \mbox{dom} \, F^*$
and $F(z_0) + F^*(z_0) = \|z_0\|^2$.  
\item For $z_0 \in \mbox{dom} \, F^*$ and $y \in \partial F^*(z_0)$
we have $F(y) + F^*(z_0) = \lan y,z_0\ran$ \cite[Thm.\ VI.5.3(c),(d)]{Ellis}.
In particular, if $y = z_0$, then $F(z_0) + F^*(z_0) = \|z_0\|^2$.  
\end{enumerate}

We first prove part (a), which is a special case of Theorem C.1 in \cite{EE}.
Let $M = \sup_{z \in \rsigma}\{F(z) - \|z\|^2/2\}$.  
Since for any $z \in \mbox{dom} \, F^*$ and $x$ in $\rsigma$
\[
F^*(z) + M \geq \lan x,z \ran - F(x) + M \geq \lan x,z \ran - \|x\|^2/2,
\]
we have
\[
F^*(z) + M \geq \sup_{x \in \rsigma} \{\lan x,z \ran - \|x\|^2/2\} = \|z\|^2/2.
\]
It follows that $M \geq \|z\|^2/2 - F^*(z)$ and thus that 
$M \geq \sup_{z \in \mbox{{\scriptsize \rm dom}} \, F^*}\{\|z\|^2/2 - F^*(z)\}$.
To prove the reverse inequality, let $N = \sup_{z \in \mbox{{\scriptsize \rm dom}} \, F^*}
\{\|z\|^2/2 - F^*(z)\}$.
Then for any $z \in \rsigma$ and $x \in \mbox{dom} \, F^*$
\[
\|z\|^2/2 + N \geq \lan x,z \ran - \|x\|^2/2 + N \geq \lan x,z \ran - F^*(x).
\]
Since $F^*(x) = \infty$ for $x \not \in \mbox{dom} \, F^*$, it follows from property 1 that
\[
\|z\|^2/2 + N \geq \sup_{x \in \mbox{{\scriptsize \rm dom}} \, F^*}\{\lan x,z \ran - F^*(x)\} = F(z)
\]
and thus that $N \geq \sup_{z \in \rsigma}\{F(z) - \|z\|^2/2\}$.

In order to prove parts (b) and (c) of Theorem \ref{thm:relatedmin},
let $z_0$ be any point in $\R^\sigma$ at which $F(z)-\frac{1}{2}\|z\|^2$ attains its
supremum.  Then $z_0 = \nabla F(z_0)$, and so by the last line of
property 2, $z_0 \in \mbox{dom} \, F^*$ and $F(z_0) + F^*(z_0) = \|z_0\|^2$.  
Part (a) now implies that
\beas
\sup_{z \in \R^\sigma} \{F(z)-\ts\frac{1}{2}\|z\|^2 \} & = & F(z_0) - \ts\frac{1}{2}\|z_0\|^2
\\
& = & \ts\frac{1}{2}\|z_0\|^2 - F^*(z_0) = \displaystyle\sup_{z \in \mbox{{\scriptsize \rm dom}}
\, F^*} \{\ts \frac{1}{2}\|z\|^2 -F^*(z)\}.
\eeas
We conclude that $\frac{1}{2}\|z\|^2  - F^*(z)$ attains its supremum on 
$\mbox{dom} \, F^*$ at $z_0$.  Not only have we proved part (b), but also we 
have proved half of part (c); namely, any global maximizer of $F(z)
- \frac{1}{2}\|z\|^2$ is a global maximizer of $\frac{1}{2}\|z\|^2 - F^*(z)$.

Now let $z_0$ be any point at which $\frac{1}{2}\|z\|^2  - F^*(z)$ attains
its supremum.  Then for any $z \in \R^\sigma$
\[
\ts \frac{1}{2}\langle z_0, z_0 \rangle - F^*(z_0) \geq \ts \frac{1}{2}\langle z, z
\rangle - F^*(z).
\]
It follows that for any $z \in \R^\sigma$
\[
F^*(z) \geq F^*(z_0) + \ts \frac{1}{2}(\lan z, z \ran - \lan z_0, z_0 \ran)  
\geq F^*(z_0) + \langle z_0, z-z_0 \rangle
\]
and thus that $z_0 \in \partial F^*(z_0)$. 
By the last line of property 3 this implies that 
$F(z_0)+F^{*}(z_0)= \|z_0\|^2$.
In conjunction with part (a) this in turn implies
that
\beas
\sup_{z \in \mbox{{\scriptsize dom}} \, F^*}  \{\ts \frac{1}{2}\|z\|^2 -F^*(z)\} & = & 
\ts \frac{1}{2}\|z_0\|^2 - F^*(z_0) \\
& = &  F(z_0) - \ts \frac{1}{2}\|z_0\|^2 = \displaystyle\sup_{z \in \R^\sigma}
\{\ts F(z)-\frac{1}{2}\|z\|^2\}.
\eeas
We conclude that $F(z)-\frac{1}{2}\|z\|^2$ attains its supremum at $z_0$.  This
completes the proof of the theorem. \ \ \ \ink

\section{Form of $\ebeta$}
\beginsec

We first derive the
form of $\ebeta$ for $\beta > 0$ as given in part (b)
of Theorem \ref{thm:eb}.  We then prove that $\ebeta = \{\rho\}$
for all $\beta \leq 0$.

$\ebeta$ is defined
as the set of $\nu \in \cp$ that minimize $R(\nu|\rho) -
\frac{\beta}{2}\lan \nu,\nu \ran$.  Since $\beta > 0$, this
is equivalent to
\be
\label{eqn:rewriteebeta}
\ebeta = \left\{\nu \in \cp : \nu \mbox{ maximizes } \ts \frac{1}{2} \lan \nu,\nu \ran
- \ts\frac{1}{\beta}R(\nu|\rho) \right\}.
\ee
This maximization problem
has the form of the right hand side of part (a) of Theorem \ref{thm:relatedmin};
viz.,
\[
\sup_{\nu \in \cp} \left\{\ts \frac{1}{2} \lan \nu,\nu \ran 
- \ts \frac{1}{\beta}R(\nu|\rho) \right\} = \sup_{\nu \in \mbox{\scriptsize dom} \, F^*}
\{\ts \frac{1}{2} \|\nu\|^2 - F^*(\nu)\}
\]
with $F^*(\nu) = \frac{1}{\beta}R(\nu|\rho)$.  
For $z \in \rq$ we define the finite, convex, continuous function
\be
\label{eqn:gammaz}
\Gamma(z) = \ts \log \left(\sum_{i=1}^q e^{z_i} \, \frac{1}{q}\right).
\ee
Since for $\nu \in \rq$ \cite[Thm.\ VIII.2.2]{Ellis}
\[
(\Gamma)^*(\nu) = \left\{ \begin{array}{ll} R(\nu|\rho) & \: \mbox{ for } \nu \in \cp \\
\infty & \: \mbox{ otherwise },
\end{array} \right.
\]
it follows that for $z \in \rq$
\[
F(z) = \sup_{\nu \in \cp}\left\{\lan z,\nu \ran - 
\ts \frac{1}{\beta}R(\nu|\rho)\right\} = \ts \frac{1}{\beta}
\ds \sup_{\nu \in \cp}\left\{\lan \beta z,\nu \ran - R(\nu|\rho)\right\} = \ts 
\frac{1}{\beta}\Gamma(\beta z).
\]
Thus by part (a) of Theorem \ref{thm:relatedmin}
\[
\sup_{z \in \rq}\left\{\ts \frac{1}{\beta} \Gamma(\beta z) - \ts \frac{1}{2}\|z\|^2
\right\} = \sup_{\nu \in \cp} \left\{\ts \frac{1}{2} \lan \nu,\nu \ran 
- \ts \frac{1}{\beta}R(\nu|\rho) \right\},
\]
and by part (b) of the theorem the global maximum points of the two functions coincide.

Equation (\ref{eqn:rewriteebeta}) now implies that
\beas
\ebeta & = & \left\{z \in \rq : z \mbox{ maximizes } 
\ts \frac{1}{\beta} \Gamma(\beta z) - \ts \frac{1}{2}\|z\|^2\right\} \\
& = & \left\{z \in \rq : z \mbox{ minimizes } 
\ts \frac{\beta}{2}\|z\|^2 - \Gamma(\beta z) \right\}.
\eeas
We summarize this discussion in the following corollary.  Part (b) of
the corollary is proved in part (b) of Theorem 2.1 in \cite{EW1}.

\begin{cor}\per
\label{cor:gbeta}
We define the finite, convex, continuous function $\Gamma$ in
{\em (\ref{eqn:gammaz})}.  The following conclusions hold.

{\em (a)} $\ebeta$ coincides with the set of global minimum points of
\[
G_\beta (z) = \ts \frac{\beta}{2} \|z\|^2 - \ds \log
\sum_{i=1}^{q}  e^{\beta z_i} = \ts \frac{\beta}{2}\|z\|^2 - \Gamma(\beta z) - \log q.
\]

{\em (b)} For $0 < \beta < \beta_c$, $\beta = \beta_c$, and $\beta > \beta_c$
the set of global minimum points of $G_\beta$ 
has the form given by the right hand side of {\em (\ref{eqn:ebetaexplicit})}
{\em [}Thm.\ {\em \ref{thm:eb}(b)}{\em ]}.
\end{cor}

Corollary \ref{cor:gbeta} completes the proof of Theorem \ref{thm:eb}.  
Michael Kiessling's proof of this corollary based on Lagrange multipliers 
is given in Appendix B of \cite{EW2}.  Continuous analogues
of the corollary are mentioned in \cite{Kie}, \cite{KieLeb}, and \cite{MesSpo}, but 
are not proved there.

We now show that for all $\beta \leq 0$, $\ebeta = \{\rho\}$.
This is obvious for $\beta = 0$ since
$\nu = \rho$ is the unique vector in $\cp$ that
minimizes $R(\nu|\rho)$.  Our goal is to prove that for 
$\beta < 0$, $\nu = \rho$ is also the unique vector in
$\cp$ that minimizes $R(\nu|\rho) - \frac{\beta}{2} \lan \nu,\nu \ran$.
Let $\bnu$ be a point in $\cp$ at which 
$R(\nu|\rho) - \frac{\beta}{2} \lan \nu,\nu \ran$ attains its infimum.
For any $i = 1,2,\ldots,q$, 
\[
\frac{\partial (R(\nu|\rho) - \frac{\beta}{2} \lan \nu,\nu \ran)}{\partial \nu_i}
= \log \nu_i + 1 - \beta \nu_i,
\]
which is negative for all sufficiently small $\nu_j > 0$.  It follows
that $\bnu$ does not lie on the relative boundary of $\cp$; i.e.,
$\bnu_j > 0$ for all $i = 1,2,\ldots,q$.  We complete the proof by
showing that for any $1 \leq j < k \leq q$, $\bnu_j = \bnu_k$.  
Since $\rho$ is the only point in $\cp$ satisfying these equalities,
we will be done.  

Given $a \in (0,1)$, we consider the reduced two-variable problem of
minimizing $R(\nu|\rho) - \frac{\beta}{2} \lan \nu,\nu \ran$ 
over $\nu_j > 0$ and $\nu_k > 0$ under the constraint
$\nu_j+\nu_k=a$; all the other components $\nu_i$ are
fixed and equal $\bnu_i$.  Setting
$\nu_k=a-\nu_j$, we define 
\[
F(\nu_j) = R(\nu | \rho) - \ts\frac{\beta}{2} \lan \nu, \nu \ran.
\]
Differentiating with respect to $\nu_j$ shows that any global
minimizer $\nu_j$ must satisfy 
\[
F'(\nu_j) = \log \nu_j - \log (a-\nu_j) - \beta(2\nu_j-a) =0.
\]
Since
\[
F''(\nu_j) = \ts\frac{1}{\nu_j} + \ts\frac{1}{a-\nu_j} - 2\beta > 0,
\]
$F'(\nu_j)$ is strictly increasing from negative
values for all $\nu_j$ near $0$ to positive values
for all $\nu_j$ near $a$.  It follows that the only root of 
$F'(\nu_j) = 0$ is $\nu_j=\frac{a}{2}$ and thus that $\nu_k = \frac{a}{2} = \nu_j$.  
Being a global minimizer of $R(\nu|\rho) - \frac{\beta}{2} \lan \nu,\nu \ran$
over $\cp$, $\bnu$ is also a global minimizer of the reduced two-variable
problem.  Since $a \in (0,1)$ is arbitrary, it follows that for any 
distinct pair of indices $\bnu_j = \bnu_k$.  This completes the proof.

\begin{acknowledgments}
The research of Marius Costeniuc and Richard S.\ Ellis
 was supported by a grant from the National Science Foundation (NSF-DMS-0202309).
The research of Hugo Touchette was supported by
the Natural Sciences and Engineering Research Council of Canada and the Royal
Society of London (Canada-UK Millennium Fellowship).
\end{acknowledgments}


\begin{thebibliography}{99}

\bibitem{Bali}
R.~Balian.
\newblock {\em From Microphysics to Macrophysics: Methods and Applications of
  Statistical Physics}, volume~I.
\newblock Springer-Verlag, Berlin, 1991.
\newblock Trans. by D. ter Haar and J. F. Gregg.

\bibitem{BarMukRuf1}
J.~Barr{\'e}, D.~Mukamel, and S.~Ruffo.
\newblock Inequivalence of ensembles in a system with long-range interactions.
\newblock {\em Phys. Rev. Lett.}, 87:030601, 2001.

\bibitem{BarMukRuf2}
J.~Barr{\'e}, D.~Mukamel, and S.~Ruffo.
\newblock Ensemble inequivalence in mean-field models of magnetism.
\newblock In T.~Dauxois, S.~Ruffo, E.~Arimondo, and M.~Wilkens, editors, {\em
  Dynamics and Thermodynamics of Systems with Long Range Interactions}, volume
  602 of {\em Lecture Notes in Physics}, pages 45--67, New York, 2002.
  Springer-Verlag.

\bibitem{BisCha}
M.~Biskup and L.~Chayes.
\newblock Rigorous analysis of discontinuous phase transitions via mean-field
  bounds.
\newblock Technical report, UCLA, 2004.
\newblock Submitted for publication.

\bibitem{BorTsa}
E.~P. Borges and C.~Tsallis.
\newblock Negative specific heat in a {L}ennard-{J}ones-like gas with
  long-range interactions.
\newblock {\em Physica A}, 305:148--151, 2002.

\bibitem{CagLioMarPul1}
E.~Caglioti, P.~L. Lions, C.~Marchioro, and M.~Pulvirenti.
\newblock A special class of stationary flows for two dimensional {E}uler
  equations: a statistical mechanics description.
\newblock {\em Commun. Math. Phys.}, 143:501--525, 1992.

\bibitem{CH1}
M.~S.~S.~Challa and J.~H.~Hetherington.
\newblock  Gaussian ensemble: an alternate Monte-Carlo scheme. 
\newblock {\it Phys.\ Rev. A} 38:6324--6337, 1988.

\bibitem{CH2}
M.~S.~S.~Challa and J.~H.~Hetherington.
\newblock  Gaussian ensemble as an interpolating ensemble. 
\newblock {\it Phys.\ Rev. Lett.} 60:77--80, 1988.

\bibitem{CosEllTou}
M.~Costeniuc, R.~S. Ellis, and H.~Touchette.
\newblock The Gaussian ensemble and universal 
ensemble equivalence for the {C}urie-{W}eiss-{P}otts model.
\newblock In preparation, 2004.

\bibitem{CosEllTouTur}
M.~Costeniuc, R.~S.~Ellis, H.~Touchette, and B.~Turkington.
\newblock The generalized canonical ensemble and its universal equivalence
with the microcanonical ensemble.  
\newblock Submitted for publication, 2004.  LANL archive: cond-mat/0408681.

\bibitem{DauHolRuf}
T.~Dauxois, P.~Holdsworth, and S.~Ruffo.
\newblock Violation of ensemble equivalence in the antiferromagnetic mean-field
  {XY} model.
\newblock {\em Eur. Phys. J. B}, 16:659, 2000.

\bibitem{DauLatRapRufTor}
T.~Dauxois, V.~Latora, A.~Rapisarda, S.~Ruffo, and A.~Torcini.
\newblock The {H}amiltonian mean field model: from dynamics to statistical
  mechanics and back.
\newblock In T.~Dauxois, S.~Ruffo, E.~Arimondo, and M.~Wilkens, editors, {\em
  Dynamics and Thermodynamics of Systems with Long-Range Interactions}, volume
  602 of {\em Lecture Notes in Physics}, pages 458--487, New York, 2002.
  Springer-Verlag.

\bibitem{EE}
T.~Eisele and R.~S.~Ellis.
\newblock Symmetry breaking and random waves for magnetic systems on a circle.
{\it Z.\ Wahrsch.\ verw. Geb.} 63:297--348, 1983.  

\bibitem{Ellis}
R.~S.~Ellis.
\newblock {\em Entropy. Large Deviations and Statistical Mechanics}.
\newblock New York: Springer-Verlag, 1985.

\bibitem{EHT}
R.~S. Ellis, K.~Haven, and B.~Turkington.
\newblock Large deviation principles and complete
equivalence and
 nonequivalence results for pure and mixed ensembles.
\newblock {\em J.\ Stat.\ Phys.} 101:999--1064, 2000.

\bibitem{EllHavTur3}
R.~S. Ellis, K.~Haven, and B.~Turkington.
\newblock Nonequivalent statistical equilibrium ensembles and refined stability
  theorems for most probable flows.
\newblock {\em Nonlinearity}, 15:239--255, 2002.

\bibitem{EllOttTou}
R.~S.~Ellis, P.~Otto, and H.~Touchette.
\newblock Analysis of phase transitions in the mean-field Blume-Emery-Griffiths
model.  Submitted for publication, 2004. LANL archive: cond-mat/0409047.

\bibitem{EllTouTur}
R.~S.~Ellis, H.~Touchette, and B.~Turkington.  Thermodynamic
versus statistical nonequivalence of ensembles for the mean-field
Blume-Emery-Griffith model. {\it Physica A} 335:518–-538, 2004. 

\bibitem{EW1}
R.~S. Ellis and K.~Wang.
\newblock Limit theorems for the empirical vector of the Curie-Weiss-Potts
model.
\newblock {\em Stoch.\ Proc.\ Appl.} 35:59--79, 1990.

\bibitem{EW2}
R.~S. Ellis and K.~Wang.
\newblock Limit theorems for maximum likelihood estimators in the
Curie-Weiss-Potts model.
\newblock {\em Stoch.\ Proc.\ Appl.} 40:251--288, 1992.

\bibitem{EyiSpo}
G.~L. Eyink and H.~Spohn.
\newblock Negative-temperature states and large-scale, long-lived vortices in
  two-dimensional turbulence.
\newblock {\em J. Stat. Phys.}, 70:833--886, 1993.

\bibitem{Gibbs}
J.~W. Gibbs.
\newblock {\em Elementary Principles in Statistical Mechanics with Especial
  Reference to the Rational Foundation of Thermodynamics}.
\newblock Yale University Press, New Haven, 1902.
\newblock Reprinted by Dover, New York, 1960.

\bibitem{Gross1}
D.~H.~E. Gross.
\newblock Microcanonical thermodynamics and statistical fragmentation of
  dissipative systems: the topological structure of the $n$-body phase space.
\newblock {\em Phys. Rep.}, 279:119--202, 1997.

\bibitem{Gross2}
D.~H.~E. Gross.
\newblock Phase transitions without thermodynamic limit.
\newblock In X.~Campi, J.~P. Blaizot, and M.~Ploszaiczak, editors, {\em
  Proceedings of Les Houches Workshop on Nuclear Matter in Different Phases and
  Transitions, Les Houches, France, 31.3-10.4.98}, pages 31--42. Kluwer Acad.
  Publ., 1999.

\bibitem{HerThi}
P.~Hertel and W.~Thirring.
\newblock A soluble model for a system with negative specific heat.
\newblock {\em Ann. Phys. (NY)}, 63:520, 1971.

\bibitem{Heth} 
J.~H.~Hetherington.
\newblock Solid $^3$He magnetism in the classical approximation.
\newblock{\it J. Low Temp. Phys.} 66:145--154, 1987.

\bibitem{HethStump}
J.~H.~Hetherington and D.~R.~Stump.
\newblock Sampling a Gaussian energy distribution to study 
phase transitions of the Z(2) and U(1) lattice gauge theories.  
\newblock{\it Phys. Rev. D} 35:1972--1978, 1987.

\bibitem{Huang}
K.~Huang.  
\newblock {\it Statistical Physics.}  
\newblock Wiley: New York, 1987.

\bibitem{IspCoh}
I.~Ispolatov and E.~G.~D. Cohen.
\newblock On first-order phase transitions in microcanonical and canonical
  non-extensive systems.
\newblock {\em Physica A}, 295:475--487, 2000.

\bibitem{JPV}
R.~S.~Johal, A.~Planes, and E.~Vives.
\newblock  Statistical mechanics in the extended Gaussian ensemble. 
\newblock {\it Phys.\ Rev.\ E} 68:056113, 2003.

\bibitem{KS}
H.~Kesten and R.~H.~Schonmann.
\newblock Behavior in large dimensions of the Potts and Heisenberg models.
\newblock {\it Rev.\ Math.\ Phys.} 1:147--182, 1990.

\bibitem{Kie}
M.~K.-H.~Kiessling.  On the equilibrium statistical mechanics of
isothermal classical self-gravitating matter.
\newblock {\it J.\ Stat.\ Phys.} 55:203--257, 1989.

\bibitem{KieLeb} M.~K.-H.~Kiessling and J.~L.~Lebowitz.
\newblock The micro-canonical point vortex ensemble: beyond equivalence. 
\newblock {\it Lett.\ Math.\ Phys.} 42:43--56, 1997.

\bibitem{KieNeu2}
M.~K.-H. Kiessling and T.~Neukirch.
\newblock Negative specific heat of a magnetically self-confined plasma torus.
\newblock {\em Proc. Natl. Acad. Sci. USA}, 100:1510--1514, 2003.

\bibitem{LanLif2}
L.~D. Landau and E.~M. Lifshitz.
\newblock {\em Statistical Physics}, volume~5 of {\em Landau and Lifshitz
  Course of Theoretical Physics}.
\newblock Butterworth Heinemann, Oxford, third edition, 1991.

\bibitem{LatRapTsa2}
V.~Latora, A.~Rapisarda, and C.~Tsallis.
\newblock Non-{G}aussian equilibrium in a long-range {H}amiltonian system.
\newblock {\em Phys. Rev. E}, 64:056134, 2001.

\bibitem{LynBelWoo}
D.~Lynden-Bell and R.~Wood.
\newblock The gravo-thermal catastrophe in isothermal spheres and the onset of
  red-giant structure for stellar systems.
\newblock {\em Mon. Notic. Roy. Astron. Soc.}, 138:495, 1968.

\bibitem{MesSpo}
J.~Messer and H.~Spohn.  Statistical mechanics of the
isothermal Lane-Emden equation.  
\newblock {\it J.\ Stat.\ Phys.}  29:561--578, 1982.

\bibitem{PeaGri}
P.~A.~Pearce and R.~B.~Griffiths.
\newblock Potts model in the many-component limit.
\newblock {\it J.\ Phys.\ A: Math.\ Gen.} 13:2143--2148, 1980.
 
\bibitem{Potts}
R.~B.~Potts.
\newblock Some generalized order-disorder transformations.
\newblock {\it Proc.\ Cambridge Philos.\ Soc.} 48:106--109, 1952.

\bibitem{Reif}
F.~Reif.  
\newblock {\it Fundamentals of Statistical and Thermal Physics}.
New York: McGraw-Hill, 1965.

\bibitem{RobSom}
R.~Robert and J.~Sommeria.
\newblock Statistical equilibrium states for two-dimensional flows.
\newblock {\em J. Fluid Mech.}, 229:291--310, 1991.

\bibitem{Rock}
R.~T.~Rockafellar.
\newblock {\it Convex Analysis}.  Princeton, NJ: Princeton Univ.\ Press, 1970.

\bibitem{Sal}
R.~Salmon.
\newblock {\em Lectures on Geophysical Fluid Dynamics}.
\newblock New York: Oxford Univ.\ Press, 1998.

\bibitem{SmiOne}
R.~A. Smith and T.~M. O'Neil.
\newblock Nonaxisymmetric thermal equilibria of a cylindrically bounded guiding
  center plasma or discrete vortex system.
\newblock {\em Phys. Fluids B}, 2:2961--2975, 1990.

\bibitem{StumpHeth} 
D.~R.~Stump and J.~H.~Hetherington.
\newblock{Remarks on the use of a microcanonical ensemble to 
study phase transitions in the lattice gauge theory}.  
\newblock{\it Phys. Lett. B} 188:359--363, 1987.

\bibitem{Thi2}
W.~Thirring.
\newblock Systems with negative specific heat.
\newblock {\em Z. Physik}, 235:339--352, 1970.

\bibitem{TouEllTur}
H.~Touchette, R.~S. Ellis, and B.~Turkington.
\newblock An introduction to the thermodynamic and macrostate levels of
  nonequivalent ensembles.
\newblock {\em Physica A}, 340:138--146, 2004.

\bibitem{Wu}
F.~Y.~Wu.  
\newblock The Potts model.  {\it Rev.\ Mod.\ Phys.} 54:235--268, 1982. 
\end{thebibliography}
\end{document}